\documentclass[prd,superscriptaddress,amsfonts,amssymb,amsmath,twocolumn,floatfix]{revtex4-2}
\usepackage{bm}
\usepackage{amsfonts}
\usepackage{latexsym}
\usepackage{graphicx}
\usepackage{amsmath}
\usepackage{palatino}
\usepackage{bigints}
\usepackage{mathpazo}
\usepackage{textcomp}
\usepackage{float}
\usepackage{booktabs}
\usepackage{dcolumn}
\usepackage{booktabs}
\usepackage{multirow}
\usepackage{hyperref}
\hypersetup{colorlinks,citecolor=blue}
\usepackage{amsmath}
\usepackage{xcolor}
\usepackage{orcidlink}
\usepackage{commath}
\usepackage{subcaption}

\def\jnl@style{\it}
\def\aaref@jnl#1{{\jnl@style#1}}

\def\aaref@jnl#1{{\jnl@style#1}}

\def\aj{\aaref@jnl{AJ}}                   
\def\apj{\aaref@jnl{ApJ}}                 
\def\apjl{\aaref@jnl{ApJ}}                
\def\apjs{\aaref@jnl{ApJS}}               
\def\apss{\aaref@jnl{Ap\&SS}}             
\def\aap{\aaref@jnl{A\&A}}                
\def\aapr{\aaref@jnl{A\&A~Rev.}}          
\def\aaps{\aaref@jnl{A\&AS}}              
\def\mnras{\aaref@jnl{Mon.~Not.~Roy.~Astron.~Soc.}}             
\def\prd{\aaref@jnl{Phys.~Rev.~D}}        
\def\prc{\aaref@jnl{Phys.~Rev.~C}}  
\def\prl{\aaref@jnl{Phys.~Rev.~Lett.}}    
\def\qjras{\aaref@jnl{QJRAS}}             
\def\skytel{\aaref@jnl{S\&T}}             
\def\ssr{\aaref@jnl{Space~Sci.~Rev.}}     
\def\zap{\aaref@jnl{ZAp}}                 
\def\nat{\aaref@jnl{Nature}}              
\def\aplett{\aaref@jnl{Astrophys.~Lett.}} 
\def\apspr{\aaref@jnl{Astrophys.~Space~Phys.~Res.}} 
\def\physrep{\aaref@jnl{Phys.~Rep.}}      
\def\physscr{\aaref@jnl{Phys.~Scr}}       
\def\commat{\aaref@jnl{Comm.~Math.~Phys.}}              
\def\science{\aaref@jnl{Science}}               
\def\cqg{\aaref@jnl{Classical Quant.~Grav.}}            
\def\jpcs{\aaref@jnl{JPCS}}                                     
\def\ijmpd{\aaref@jnl{Int.~J.~Mod.~Phys.~D}}                    
\def\grg{\aaref@jnl{Gen.~Relat.~Gravit.}}               
\def\rpp{\aaref@jnl{Rep.~Prog.~Phys.}}          
\def\npa{\aaref@jnl{Nucl.~Phys.~A}}        
\def\lrr{\aaref@jnl{Living Rev.~Rel.}}                   
\def\jcap{\aaref@jnl{J.~Cosmology Astropart.~Phys.}}    
\def\rmp{\aaref@jnl{Rev.~Mod.~Phys.}}   
\def\epjc{\aaref@jnl{Eur.~Phys.~J.~C}}

\allowdisplaybreaks[1]

\begin{document}

\color{black}       

\title{Phase structures and critical behaviour of rational non-linear electrodynamics AdS black holes in Rastall gravity}

\author{Yassine Sekhmani\orcidlink{0000-0001-7448-4579}}
\email[Email: ]{sekhmaniyassine@gmail.com}
\affiliation{Ratbay Myrzakulov Eurasian International Centre for Theoretical
Physics, Astana 010009, Kazakhstan.}
\affiliation{L. N. Gumilyov Eurasian National University, Astana 010008,
Kazakhstan.}
\author{Dhruba Jyoti Gogoi\orcidlink{0000-0002-4776-8506}}
\email[Email: ]{moloydhruba@yahoo.in}
\affiliation{Department of Physics, Moran College, Moranhat, Charaideo 785670, Assam, India.}
\affiliation{Theoretical Physics Division, Centre for Atmospheric Studies, Dibrugarh University, Dibrugarh
786004, Assam, India.}
\author{Ratbay Myrzakulov \orcidlink{0000-0002-5274-0815}}
\email[Email: ]{rmyrzakulov@gmail.com }
\affiliation{L. N. Gumilyov Eurasian National University, Astana 010008,
Kazakhstan.}
\affiliation{Ratbay Myrzakulov Eurasian International Centre for Theoretical
Physics, Astana 010009, Kazakhstan.}
\author{ Javlon Rayimbaev\orcidlink{0000-0001-9293-1838}}
\email[Email: ]{javlon@astrin.uz}
\affiliation{
New Uzbekistan University, Movarounnahr Str. 1, Tashkent 100000, Uzbekistan}
\affiliation{
University of Tashkent for Applied Sciences, Gavhar Str. 1, Tashkent 100149, Uzbekistan }
\affiliation{National University of Uzbekistan, Tashkent 100174, Uzbekistan}

\begin{abstract}

This research paper presents a black hole solution with a rational non-linear electrodynamics source in the Rastall gravity framework.  The paper analyzes the thermodynamic properties of the solution in normal phase space and explores its critical behavior. The phase structure is examined using the extended first law of thermodynamics, with the cosmological constant $\Lambda$ serving as pressure $P$. The isotherms exhibit van der Waals behavior at small values of horizon $r_+$. The paper also investigates the Gibbs free energy behavior and finds two critical points with two pressures where the reentrant phase transition occurs and disappears. { On the other hand, we explore the prevalent microstructure of black holes in Ruppeiner geometry, uncovering significant deviations in the nature of particle interactions from conventional practice. Moreover,} the thermodynamic geometry is analyzed using the Ruppeiner formalism, with the normalized Ricci scalar indicating possible point-phase transitions of the heat capacity, and the normalized extrinsic curvature having the same sign as the normalized Ricci scalar. The three-phase transitions of the heat capacity are those that we find for the normalized Ruppeiner curvatures. Thus, there is an absolute correspondence.

\end{abstract}


\maketitle

\section{Introduction}
The theory of general relativity (GR) is the best-known initial theory of the geometric framework for gravity \cite{1916_Einstein}. Within each of the predictions, the presence of black holes and Gravitational waves (GWs) is the most enthralling prediction of GR. About a century after its prediction, the feasibility of GR gained traction with direct observation of GWs by the twin detectors of the Laser Interferometer Gravitational Wave Observatory (LIGO) \cite{2016_Abbott}. These ground-based LIGO, Variability of Solar Irradiance, and Gravity Oscillations (Virgo) detector systems anticipated the binary black hole systems, which served as a credible proof for the validity of GR. Furthermore, the direct detection of GWs opened up a new pathway for testing or constraining alternative or modified gravity theories, as well as GR. In weak field regimes as well as moderately relativistic regimes, GR is shown to pass several physical tests, such as solar system tests \cite{will2014} and constraints from binary pulsars \cite{hulse1975, damour1992}. Furthermore, the discovery of GWs by LIGO and Virgo detectors has confirmed their validity in the extremely relativistic strong gravity regime associated with binary black holes. Hence, these GWs have significantly altered our picture of the cosmos. Recent studies show that the properties of GWs might vary based on the type of modified gravity. For example, in metric f(R) gravity, GWs possess three polarization modes: tensor plus and cross or GR modes, and scalar polarization mode. The tensor plus and cross modes propagate through spacetime at the speed of light and are, in general, transverse, traceless, and massless in nature. The scalar polarization mode, on the other hand, is a hybrid of the massless breathing mode and the massive longitudinal mode \cite{Liang_2017, gogoi1, gogoi2}. The massless breathing mode is transverse in nature but not traceless. Consequently, the investigation of the properties of black holes (which are regarded as the cleanest objects in the universe) and other compact stars like neutron stars (potential candidates for the generation of GWs) in the modified gravity regime was prompted by these new and fascinating characteristics of GWs in different modified theories of gravity. 

The $f(R)$ theories are commonly considered as one of the simplest extensions of GR so far. Apart from $f(R)$, another simple extension with violation of energy-momentum conservation is Rastall gravity, which was introduced in 1972 by P. Rastall \cite{1972_Rastall}. This idea attracted little attention to the scientific community when it was first proposed. But recently it has attracted a large number of researchers due to its simple field equations and unique black hole solutions. 
The original conservation law is modified in this theory of gravity by setting the covariant divergence of the energy-momentum tensor proportional to the covariant divergence of the Ricci curvature scalar. Regardless, by merely setting the background curvature to zero, the typical conservation rule of GR can simply be obtained. It should also be noted that, in the absence of any matter source, the Rastall gravity is equivalent to the GR. This modified gravity has recently been employed in a diversity of contexts. Many contributors, \cite{Oliveira,Heydarzade}, have already studied black holes and neutron stars in the framework of this theory of gravity. Static and spherically symmetric solutions for neutron stars are discussed in Ref. \cite{Oliveira}. In that theory, too, the work done on the solutions of black holes is given in Ref. \cite{Heydarzade}. A study on the possible solutions of black holes surrounded by ideal fluids is presented in another publication of the same year, \cite{Heydarzade2}. Quasi-normal modes are studied in \cite{Liang} for black holes surrounded by quintessence fields in Rastall gravity. This work has revealed the appropriate quasi-normal modes as a function of the parameters of the Rastall gravity model. In this work, two cases are taken into discussion: one refers to $\kappa\,\lambda<0$, which exhibit faster damping for gravitational, electromagnetic, and massless scalar fields. The second one deals with $\kappa\,\lambda>0$, from which slower damping is present for gravitational, electromagnetic, and massless scalar fields. In the former scenario, the actual frequencies of the quasi-normal modes were found to be higher than in the GR. On the contrary, in the second scenario, they were lower. Also, they deduced that the changes in the frequencies of the real and imaginary quasi-normal modes with $\kappa \lambda$ are comparable for different values of $l$ and $n$ \cite{Xu, Lin, Hu}. In parallel, many attempts have been made to reveal the phase structure of black holes in the framework of Rastall gravity. In particular, according to \cite{Rasfl}, a charged AdS black hole surrounded by a perfect fluid is constructed within Rastall gravity. Indeed, the matter sector can be characterized in terms of multiple probable kinds of rational non-linear electrodynamics sources, such as dust and radiation, or exotic matter, like a quintessence, $\Lambda$CDM type, and phantom fields. For a choice of a phantom field, the charged AdS black hole shows what is called a RPT when some of its parameter spaces satisfy certain constraints.

The performance of the critical study for revealing phase transitions is roughly a major concreting step for black hole thermodynamics. Hawking and Page demonstrated from their contribution work that there is a range of first-order black hole phase transitions between large, stable black holes and thermal radiation in anti-de Sitter space \cite{Hawking}. This work initiated the possibility of investigating the thermodynamic phase transition in black holes. An emerging reason is constructed within the anti-de Sitter/conformal field theory theories  (AdS/CFT) \cite{Maldacena,Gubser,Witten}, implying at the level of the Hawking-Page phase transition that this transition can be interpreted as the gravitational dual of the confinement/deconfinement phase transition of gauge fields \cite{Witten1}. Then, the AdS/CFT duality stretched out onto the charged and rotating AdS black holes \cite{Chamblin,Chamblin1,Caldarelli}. 

Recently, the pioneering of AdS black hole thermodynamics is widely regarded as a new version of phase space, or extended phase space, after considering the cosmological constant as thermodynamic pressure \cite{Henneaux,Teit,Kastor,Dolan,Cvetic}. Roughly speaking, this consideration led to some thermodynamic phase transition results for a class of ordinary black holes. In AdS spaces, the charged black hole reveals Van der Waals-like behavior \cite{Chamblin,Chamblin1}. The critical study for the charged black hole, or the Reissner–Nordström–AdS (RN-AdS), undergoes a first-order phase transition, revealing a similarity with the Van der Waals liquid–gas phase transition. The critical behavior of the black hole along the $P$-$V$ diagram is performed in extended phase space by Kubiznak and Mann \cite{Mann,Mann1}, which reinforced the relation between charged AdS black holes and liquid-gas Van der Waals systems \cite{Kastor,Kastor1,Dolan,Dolan1,Mann,Mann1}. { On the other hand, the emergence of thermodynamic geometry inspired by Ruppeiner formalisms is an endeavor to draw, phenomenologically or qualitatively, insights into the microscopic interactions of a set system from the axioms of thermodynamics. The key idea is that, in ordinary thermodynamic systems, the curvature of Ruppeiner metrics is linked to the nature of the interactions among the underlying particles. Particularly for systems where the microstructures interact in an attractive pattern, the curvature scalar is assigned a negative sign, versus a positive one for mostly repulsive forces. Further, the metric is flat for systems with no interaction, like the ideal gas. This approach has been successfully demonstrated for a broad variety of statistical and physical models. Excitingly, new studies have highlighted that this is viable for black holes as well, thereby supplying an empirical tool for deriving the microstructure of black holes from macroscopic insight \cite{Cai:1998ep,Wei:2015iwa,Wei:2019uqg,Guo:2019oad,Xu:2020gud,Ghosh:2020kba,Dehghani:2023yph,Sekhmani:2023plr}, notwithstanding the absence of a quantum theory of gravitation.}  

The latest activities in which the physical aspect is thermodynamics or whether (optics, quasi-normal modes,...) in the essence of a black hole spacetime in the context of modified gravity models are now constituting a large interest in theoretical physics. The purpose of examining such an extension of general relativity is considered a step toward reinforcing the general theory of relativity and establishing solid approaches to it. Nowadays, most works are taken in the essence of modifying gravity through the study of phase transitions \cite{p1,p2,p3,p4,p5,p6,p7,pq}, quasinormal modes \cite{q1,q2,q3,q4,q5,q6,q7}, and shadow behaviors for black hole spacetime \cite{s1,s2,s3,s4,s5,s6}.

The kind leading to the existence of such black holes with regular spacetime is related as a consequence of curvature singularities, implying the non-singular behavior of spacetime structure. To avoid the singularity, Gliner \cite{Gliner} and Sakarov \cite{sakharov} modeled a situation in which the matter source behaves like a de Sitter core at the center of spacetime with the state equation $p=-\rho$. As the approach to the singularity problem was clearly understood, Bardeen invented a way to find the first nonsingular solution of Einstein's equations, namely, the Bardeen regular black hole \cite{Bardeen}. The solution is nowhere singular and therefore regular in all spacetime, involving a de Sitter center, according to suggestions in Sakharov's work, as well as satisfying the weak energy condition. The possible classification of the famous nonlinear electrodynamic source is followed up with the Bardeen, Hayward, and Ayon-Beato Garcia kinds of matter sources, and thereby, in general relativity, their possible solutions to the black hole are examined suitably \cite{Hayward,ABG}. 

The outline of this work is as follows: The first section consists of building a black hole solution in Rastall gravity with a rational non-linear electrodynamics source. The second section is devoted to determining all the thermodynamic quantities in the normal phase space, such as temperature, entropy, heat capacity, and Gibbs free energy. Next, the third section is a part that takes care to study, in extended phase space, the phase structure along the P-V criticality and shows the behavior of the Gibbs free energy. The last section deals with the investigation of thermodynamic geometry. Finally, we summarize our work with a conclusion.

\section{Rastall gravity}

This section's goal is to obtain a solution while taking into account the physical background of the gravity sector as well as the matter source. To begin working toward this goal, the computation includes, in a way, two blocks regarding the model: the Rastall field equations and the matter field equations. We begin by looking at the Rastall gravity sector.

In this interesting modification of GR, the covariant conservation condition $T^{\mu \nu}_{\ ; \, \nu} = 0$ was changed to a more generalized version as:
\begin{equation}\label{r1}
\nabla_\nu T^{\mu\nu} = a^\mu.  
\end{equation}
To make the theory consistent with relativity, one must have the right-hand side of the above equation equal to zero when the scalar curvature or the background curvature vanishes. Thus, for a convenient option, we can set the four-vector $a^\mu$ as,
\begin{equation}\label{r2}
 a^\mu = \lambda \nabla^\mu R,  
\end{equation}
here $\lambda$ is a free parameter known as Rastall's parameter. From Eq.s \eqref{r1} and \eqref{r2}, we can have the Rastall field equation as:
\begin{eqnarray}\label{r3}
&& R_{\mu\nu}-\frac{1}{2}\left(\, 1 -2\beta\,\right) g_{\mu\nu}R=\kappa T_{\mu\nu}\ ,\label{E0}\\
\end{eqnarray}
where $\beta = \kappa \lambda$ and from hereafter, we shall term this as the Rastall parameter.
The trace of the above equation gives,
\begin{equation}\label{r4}
R=\frac{\kappa}{\left(4\,\beta-1\right)}\,T\ ,\quad \beta\ne 1/4.
\end{equation}

The field equation in the presence of a non-vanishing cosmological constant is
\begin{equation}\label{r5}
E_{\mu\nu}+ \Lambda g_{\mu \nu} + \beta \,g_{\mu\nu}  R  = \kappa T_{\mu\nu},
\end{equation}
where $E_{\mu\nu}$ is the standard Einstein tensor.

\section{Generalized AdS Spherically Symmetric BH Solution in Rastall’s gravity}
To better approach the given model of Rastall gravity coupled to the term of rational non-linear electrodynamics fields, we assume an ansatz for the spacetime geometry. Indeed, the spacetime metric is considered to be static, spherically, and symmetric in the following form:
\begin{equation} \label{r6}
\mathrm{d}s^2 = -f(r) \mathrm{d}t^2 + \dfrac{\mathrm{d}r^2}{f(r)} +r^2 \mathrm{d}\Omega^2,
\end{equation}
where $f(r)$ is the metric function which depends on the radial coordinate  $r$ and $\mathrm{d}\Omega^2=\mathrm{d}\theta^2+\sin^2\theta \mathrm{d}\phi^2$. According to the space-time structure, we assume a particular ansatz for the gauge field in the following form:
\begin{equation}\label{r7}
A=Q_m\cos\theta \mathrm{d}\phi
\end{equation}
where the parameter $Q_m$ stands for the total magnetic charge, and the quadratic invariant $\mathcal{F}$ is explicitly given by
\begin{equation}\label{r8}
 Q_m=\frac{1}{4\pi}\int \mathcal{F},\quad\quad   \mathcal{F}=\frac{2Q_m^2}{r^4}.
\end{equation}
The conjugate potential, on the other hand, maybe redefined as
\begin{equation}\label{r9}
\Psi=\widetilde{A}_t(r_+)-\widetilde{A}_t(\infty),\quad \widetilde{F}=\mathrm{d}\widetilde{A}=\mathcal{L}_\mathcal{F}\,\,^{\star}F.
\end{equation}
The definition of matter sector, involving a magnetic charge parameter to the black hole system, is a generalized term for both the type $(I)$ and the type $(II)$ classes of rational non-linear electrodynamics fields. By the way, according to \cite{ff}, the generic expression is perfectly expressed as
\begin{equation}\label{r10}
    \mathcal{L}(\mathcal{F})=\frac{4\mu}{\alpha}\frac{\left(\alpha\mathcal{F}\right)^{\frac{\nu+3}{4}}}{\left(1+\left(\alpha\mathcal{F}\right)^{\frac{\nu}{4}}\right)^{\frac{\mu+\nu}{\nu}}},
\end{equation}
in which $\mu>0$ is a dimensionless constant and $\alpha>0$ carries the dimension of length squared. The generic formulation of the Lagrangian is a function of the quadratic invariant term  $\mathcal{F}=F_{\mu\nu}F^{\mu\nu}$ where $F_{\mu\nu}=\partial_{[\mu}A_{\nu]}$. A closer look reveals the following related expressions in the cases: $\nu=\mu,2,1$
\begin{equation}\label{r11}
\mathcal{L}(\mathcal{F})=
    \begin{cases}
      \frac{4\mu}{\alpha}\frac{\left(\alpha\mathcal{F}\right)^{\frac{\mu+3}{4}}}{\left(1+\left(\alpha\mathcal{F}\right)^{\frac{\mu}{4}}\right)^{2}}\qquad\text{type $I$} \left(\nu=\mu\right),\\
      \frac{4\mu}{\alpha}\frac{\left(\alpha\mathcal{F}\right)^{\frac{5}{4}}}{\left(1+\left(\alpha\mathcal{F}\right)^{\frac{1}{2}}\right)^{\frac{\mu+2}{2}}}\hspace*{0.365cm}\text{type $II$} \left(\nu=2\right),\\
      \frac{4\mu}{\alpha}\frac{\left(\alpha\mathcal{F}\right)}{\left(1+\left(\alpha\mathcal{F}\right)^{\frac{1}{4}}\right)^{{\mu+1}}}\hspace*{0.365cm}\text{Maxwellian type} \left(\nu=1\right),
    \end{cases}\,.
\end{equation}
It is interesting to note that, while considering the weak field limit, new reduced expressions for the considered Lagrangian are generated in the approximative form as 
\begin{equation}\label{r12}
\mathcal{L}(\mathcal{F})\sim
    \begin{cases}
    \alpha^{\frac{\mu-1}{4}}\mathcal{F}^{\frac{\mu+3\nu}{4}}\qquad \text{type $I$},\\
       \alpha^{\frac{1}{4}}\mathcal{F}^{\frac{5}{4}}\hspace*{1.4cm} \text{type $II$},\\
       \mathcal{F}\hspace*{2.05cm} \text{Maxwellian type}.
    \end{cases}\,.
\end{equation}
Essentially, the type $(I)$ with the conditions ($\mu>1$) and ($0<\mu<1$) involves, respectively, a stronger and weaker vector field than the Maxwell vector field. Further, the case $\mu=1$ gives rise to a pure Maxwell field in the weak-field limit. Similarly, thinking about the type $(II)$ as pointed out, it might be slightly stronger than the Maxwell field. Indeed, the preceding claim demonstrates that type $(II)$ is a possible approach type of the Maxwell field in the weak field limit and is therefore the best candidate for the rational non-linear electrodynamics source of black holes.

Let us now look at the gravitational and matter-source equations of motion (EOM). We can now obtain the following non-vanishing components of the field equation by defining the Rastall tensor from the field equation Eq.~\eqref{r5} as $\Theta_{\mu\nu} = E_{\mu\nu}+\Lambda g_{\mu\nu} + \kappa \lambda g_{\mu\nu} R$ such that
\begin{eqnarray}\label{r13}
&&{\Theta^{0}}_{0}=\frac{1}{r^2}\big[rf^{\prime}(r) + f(r) -1 \big]+ \Lambda+\beta R,\nonumber\\
&&{\Theta^{1}}_{1}=\frac{1}{r^2}\big[rf^{\prime}(r) + f(r) - 1  \big]+ \Lambda+\beta R,\nonumber\\
&&{\Theta^{2}}_{2}=\frac{1}{r^2}\big[rf^{\prime}(r)+\frac{1}{2}r^2 f^{\prime\prime}(r)\big]+ \Lambda+\beta R,\nonumber\\
&&{\Theta^{3}}_{3}=\frac{1}{r^2}\big[rf^{\prime}(r)+\frac{1}{2}r^2 f^{\prime\prime}(r)\big]+ \Lambda+\beta R,
\end{eqnarray}
where the Ricci scalar reads as
\begin{equation}\label{r14}
R=-\frac {1}{{r}^{2}}\big[{r}^{2}f^{\prime\prime}(r) + 4r f^{\prime}(r) + 2\,f(r) -2 \big].
\end{equation}
Here the prime denotes the derivative with respect to the radial coordinate 
$r$. It is seen that the mixed Rastall tensor components ${\Theta^{0}}_{0} = 
{\Theta^{1}}_{1}$ and ${\Theta^{2}}_{2} = {\Theta^{3}}_{3}$. This is the 
consequence of the spherical symmetric nature of the metric (\ref{r6}) in 
the mixed tensor form. In this work, we would like to consider a general total 
energy-momentum tensor $T_{\mu\nu}$ defined by
\begin{equation} \label{r15}
T_{\mu\nu}=\frac{2}{\kappa}\left(\mathcal{L}_{\mathcal{F}}F_{\mu\nu}^2-\frac{1}{4}g_{\mu\nu}\mathcal{L}(\mathcal{F})\right).
\end{equation}
In this case, $\mathcal{L}_\mathcal{F}$ is the first derivative of the function $\mathcal{F}$. While, the antisymmetric Faraday tensor $F_{\mu\nu}$ satisfies the following nonlinear Maxwell equations 
\begin{eqnarray}\label{r16}
\partial_\mu\left(\mathcal{L}_{\mathcal{F}}F^{\mu\nu}\right)=0.
\end{eqnarray}
Proceeding with the matter part of the field equations, the energy-momentum tensor projection is now expressed as follows:
 \begin{eqnarray}\label{r17}
&&{T^{t}}_{t}={T^{r}}_{r}=-\frac{1}{2}\mathcal{L},\nonumber\\
&&{T^{\theta}}_{\theta}={T^{\phi}}_{\phi}=-\mathcal{L}+\frac{4Q_m^4}{r^4}\mathcal{L}_{\mathcal{F}}.
\end{eqnarray}
So, the now-obtained set of EOM now provides an easy way to clearly define each component of the gravitational and matter sectors. To proceed, we will demonstrate the expression of the components $(t, t)$, $(r, r)$, $(\theta, \theta)$, and $(\phi, \phi)$. Thus the ${\Theta^{t}}_{t}={T^{t}}_{t}$ and ${\Theta^{r}}_{r}={T^{r}}_{r}$ 
components of Rastall field equations give,
\begin{widetext}
\begin{equation}\label{r18}
\frac{1}{r^2}\Bigg(rf^{\prime} + f  -1 \Bigg) + \Lambda-\frac {\beta
}{{r}^{2}}\Bigg({r}^{2}f^{\prime\prime} + 4r f^{\prime}+2\,f -2\Bigg)=-\frac{2^{\frac{\nu +7}{4}}\,   \left(\frac{\alpha  Q_m^2}{r^4}\right)^{\frac{\nu +3}{4}}}{\alpha }\mu\,{\mathcal{H}^{-\frac{\mu +\nu }{\nu }}},
\end{equation} 
\end{widetext}
and ${\Theta^{2}}_{2}={T^{2}}_{2}$ and ${\Theta^{3}}_{3}={T^{3}}_{3}$ components can be read as
\begin{widetext}
\begin{equation}\label{r19}
\frac{1}{r^2}\Bigg(rf^{\prime}+\frac{1}{2}r^2 f^{\prime\prime}\Bigg) + \Lambda-\frac {\beta
}{{r}^{2}}\Bigg({r}^{2}f^{\prime\prime}  +4r f^{\prime} +2\,f  -2 \Bigg)=-\frac{2^{\frac{\nu +7}{4}} \left(\frac{\alpha  Q_m^2}{r^4}\right)^{\frac{\nu +3}{4}}}{\alpha}\mu\,\mathcal{G}\,\mathcal{H}^{-\frac{\mu }{\nu }-2}
\end{equation}
\end{widetext}
where $\mathcal{H}$ and $\mathcal{G}$ are expressed according to the terms
\begin{align}\label{r20}
\mathcal{H}&=1+2^{\nu /4}
   \left(\frac{\alpha  Q_m^2}{r^4}\right)^{\nu /4}, \\
   \mathcal{G}&=2^{\nu /4}
   \Bigg((\mu -3) Q_m^2+2\Bigg) \left(\frac{\alpha  Q_m^2}{r^4}\right)^{\nu /4}\nonumber \\
   &+2-3 Q_m^2-\nu  Q_m^2.\hspace{1cm}
\end{align}
The present step is restricted to discovering a possible solution for the metric function $f(r)$; for that reason, we use the differential set identified by the component $(t, t)$. The approach to solving the second-order differential equation for the variable function $f(r)$ in this situation produces the following modeling solution: 
\begin{widetext}
\begin{equation}\label{r21}
f(r)=
   1-\frac{2 M}{r}+\frac{\Lambda  r^2}{12 \beta -3}+\frac{2 q^3}{\alpha\, r (1 -\beta )}\left(1-\frac{r^\mu}{(r^\nu+q^\nu)^{\frac{\mu}{\nu}}}\right)
+ g(r)\,\, {_2}F_1\left(a,b;c;z\right),
\end{equation}
\end{widetext}
where 
\begin{equation}\label{r22}
    g(r)=2\,r^2 \left(\frac{q}{r}\right)^{\nu +3}\frac{\beta\,  \mu  }{\alpha  (\beta -1) (\beta  (\nu -1)+1)},\quad \beta\neq1
\end{equation}
and $M$ is an integration constant standing for the mass of the black hole and $q$ is a free integration constant related to the magnetic charge, so that  
\begin{equation}\label{r23}
    Q_m=\frac{q^2}{\sqrt{2\alpha}}.
\end{equation}
Here, the hypergeometric function ${_2}F_1\left(a,b;c;z\right)$ representing the regular solution of the hypergeometric differential equation, is defined for $\lvert z\rvert<1$  by a power series of the form ${_2}F_1\left(a,b;c;z\right)=\sum_{k=0}^{\infty}\bigg[(a)_k(b)_k/(c)_k\bigg]z^k/k!$, where $(n)_k$ is the (rising) Pochhammer symbol \cite{pon}. In fine order, the elements of the hypergeometric function employing the metric function are given as follows:
ç*àpggc \begin{equation}\label{r24}
    \begin{cases}
      a=\frac{\beta  (\nu -1)+1}{\beta  \nu },
      \vspace{3mm}\\
      b=\frac{\mu +\nu }{\nu }, \vspace{3mm}\\
     c=\frac{\frac{1}{\beta }-1}{\nu }+2, \vspace{3mm}\\
     z=-\left(\frac{q}{r}\right)^{\nu}.
    \end{cases}\,.
\end{equation}
It is worth noting that according to \cite{Ashtekar}, the AMD mass could be expressed as
\begin{equation}\label{r25}
M_{\text{ADM}}=M+\alpha^{-1}q^3.
\end{equation}
Regarding the ADM mass term, or rather the condensate of the massless graviton, a close examination shows that the mass term is divided into two natures, one of which refers to the self-interaction involving the ordinary physical mass term $M$, and the second is nothing more than the non-linear interactions between the graviton and the (nonlinear) photon, leading to a charge contribution $\alpha^{-1}q^3$.

To have a better understanding of the black hole system, it might be useful to collect all of the parameters into the parameter space. Thus, $\left(M, q, \mu, \nu, \beta, \Lambda\right)$ expresses the underlying parameter space. Resuming some limits over the parameter space will allow us to find another black hole system. The limit $\left(\beta = 0 \right)$, in particular, disables the presence of Rastall gravity, implying that the black hole is a solution in the frame of General Relativity. At such considered limits, the black hole solution behaves as \cite{ElMoumni,Hawking}
\begin{align}
   & f(r)\bigr\rvert_{\beta\rightarrow0}=1-\frac{2M}{r}-\frac{\Lambda r^2}{3}+\frac{2q^3}{\alpha\, r}\left(1-\frac{r^\mu}{(r^\nu+q^\nu)^{\frac{\mu}{\nu}}}\right)\hspace{0.3cm}\nonumber\\
    & f(r)\bigr\rvert_{\beta\rightarrow0,\,q=0}=1-\frac{2M}{r}-\frac{\Lambda r^2}{3}.\nonumber
\end{align}

To conduct some analysis of nature's singularities in relation to the physical solution $(\ref{r21})$, the Ricci square ($R_{\mu\nu}R^{\mu\nu}$) and Kretshmann scalars ($R_{\mu\nu\lambda\sigma}R^{\mu\nu\lambda\sigma}$) are needed.
\begin{widetext}
\begin{eqnarray}\label{r27}
R_{\mu\nu}R^{\mu\nu}&=&\frac{r^4 f''(r)^2+8 r^2 f'(r)^2+8 f(r) \left(r f'(r)-1\right)+4 r f'(r) \left(r^2 f''(r)-2\right)+4 f(r)^2+4}{2 r^4}.\\
R_{\mu\nu\lambda\sigma}R^{\mu\nu\lambda\sigma}&=&f''(r)^2+\frac{4 f'(r)^2}{r^2}+\frac{4 (f(r)-1)^2}{r^4}.
\end{eqnarray}
\end{widetext}
Using the metric function in our case, one can see that the scalars are not free from singularity issues. This is mainly due to the presence of the violation of energy-momentum conservation.

\begin{figure*}[htb!]
      	\centering{
      	\includegraphics[scale=0.42]{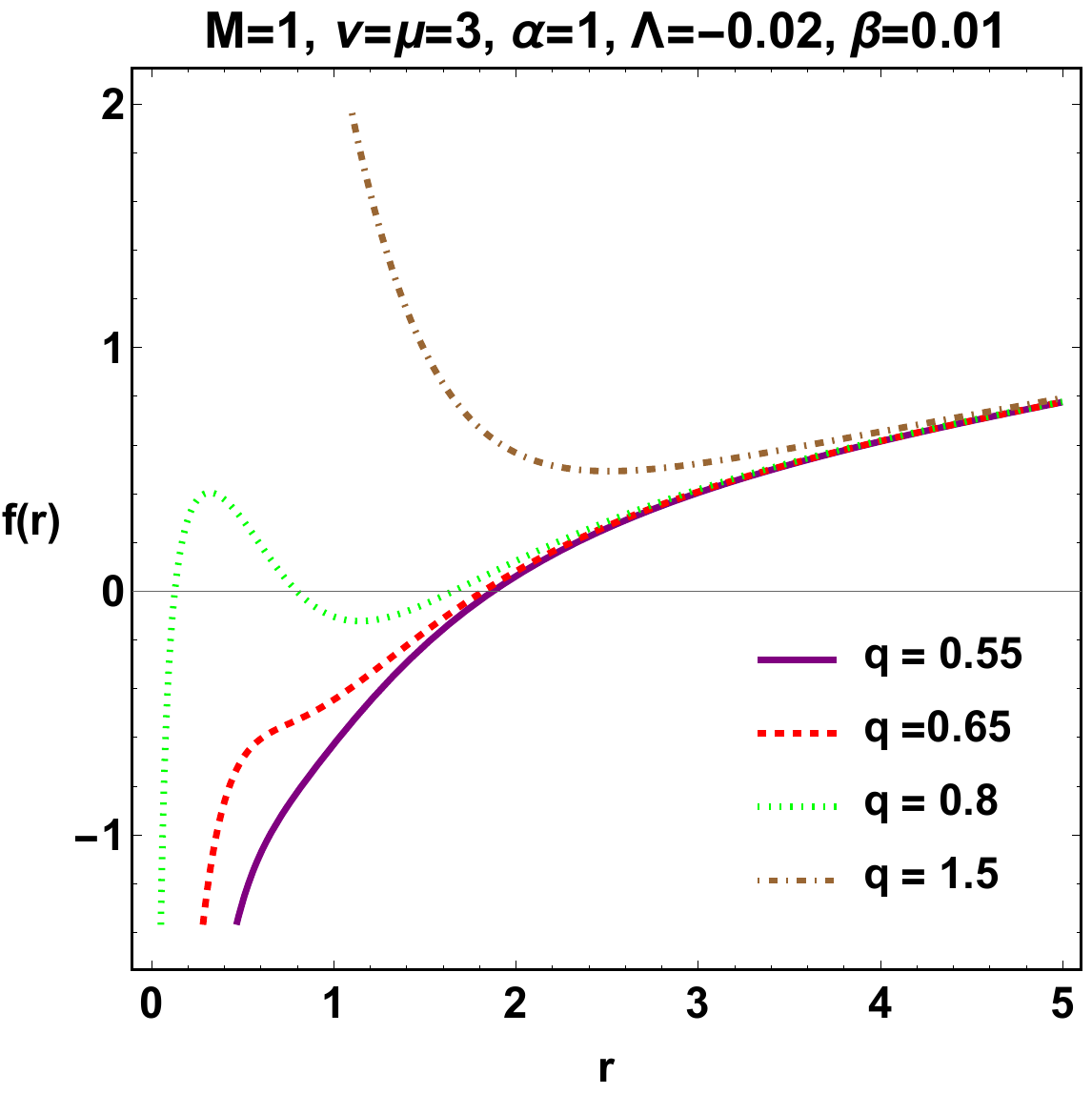} \hspace{0.2mm}
       \includegraphics[scale=0.42]{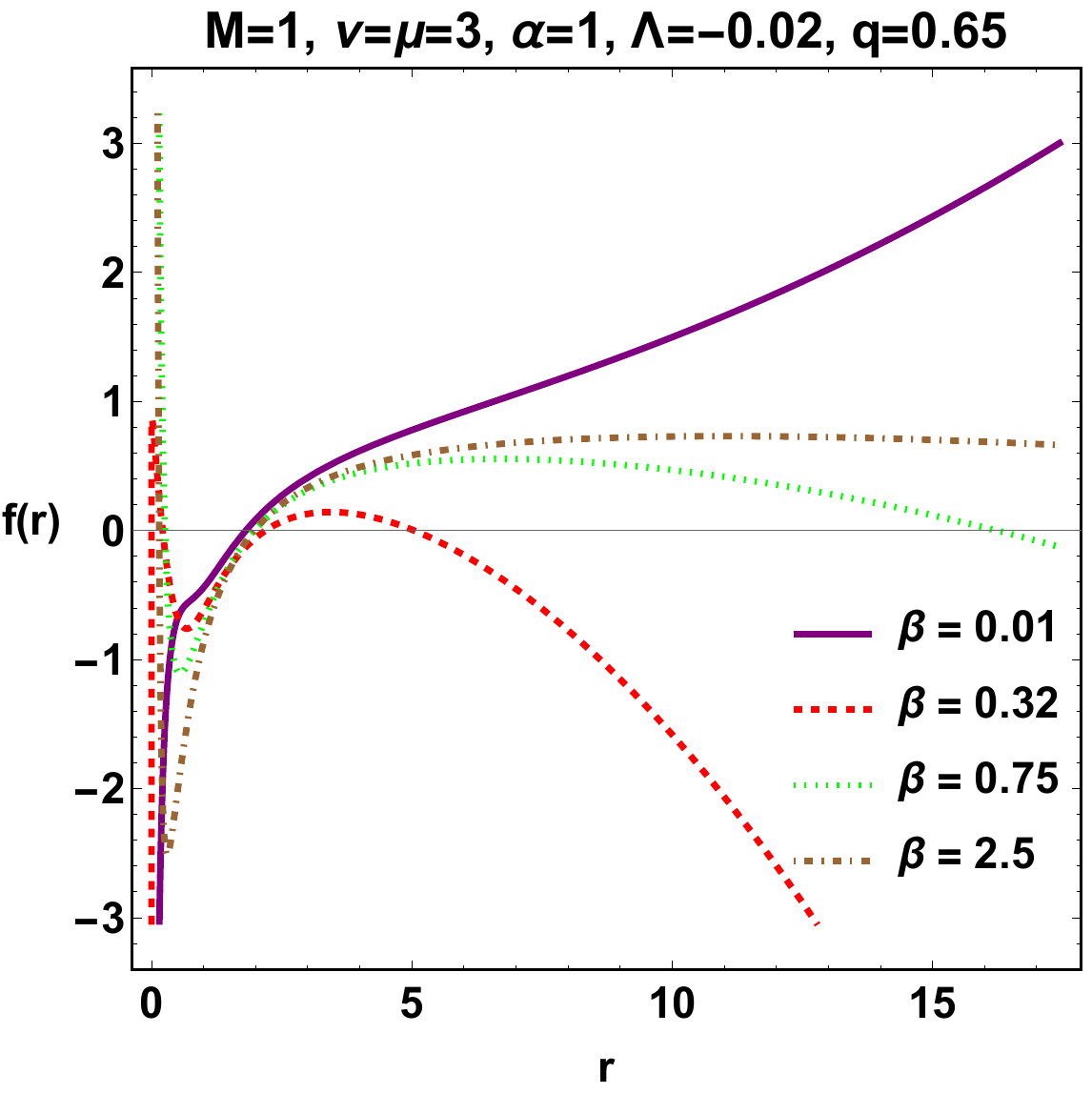}}
      	\caption{Variation of the black hole metric function (\ref{r21}) with respect to $r$ for various values of $(q, \beta)$ and for fixed values of ($M, \nu, \mu, \alpha, \Lambda, \beta$).}
      	\label{fig1}
      \end{figure*}

{ On the graphical side, the metric function is represented (Fig. \ref{fig1}) by taking into account two parameter variations, namely, $\beta$ and $q$. It is interesting to set out some critical situations concerning the associated horizon structure. In this way, the variation of the parameter $\beta$ involves some interesting behaviors. In particular, according to the value of $\beta$, the number of possible sets of the horizon radius is dependent. In other words, it is found that there exist two critical scenarios in which the number of horizon radii changes. Hence, the case $\beta =0.32$ could generate four existing horizon radii, namely, the smallest root corresponds to two black hole horizons with an extra degeneracy smallest one, and the largest root is associated with a cosmological horizon. Moreover, for the case $\beta>\beta_c =0.32$, the horizon set carries at most three or two horizon radii. However, the case $\beta<\beta_c =0.32$ could generate one possible horizon radius. On the other hand, the variation of the parameter $q$, in particular for $q = 0.8$, generates at most three horizon radii, namely two black hole horizons and the cosmological horizon. In addition, some critical issues are constrained by the critical value $q_c = 0.8$, for which the case $q<q_c $ preserved only one horizon radius; otherwise, the situation refers to the naked singularity indicating an empty for the horizon roots.}

\section{Thermodynamic properties}
To study in depth the thermodynamic properties in the normal phase space of the resulting black hole solution, it is necessary to have an overview of the thermodynamic quantities in terms of mass, Hawking temperature, heat capacity, and Gibbs free energy, i.e. $(M, T_+, C_+, F_+)$. To construct these quantities, we first look for the mass of the black hole. At the radius of the $r_+$ horizon of the metric solution (\ref{r21}), the mass of the black hole is given by
\begin{widetext}
    \begin{align}\label{r28}
   M&= \frac{1}{2} r_+ \Bigg\{1+ \frac{\Lambda  r_+^2}{12 \beta -3}+\frac{q^3
   }{\alpha  (1-\beta ) r_+}\left(2-2 \left(\left(\frac{q}{r_+}\right)^{\nu }+1\right)^{-\frac{\mu }{\nu }}\right)
   +\left(\frac{q}{r_+}\right)^{\nu +2}\frac{2 \beta\,  \mu \, q \,r_+  }{\alpha  (\beta -1) (\beta  (\nu -1)+1)}\nonumber\\
   &\times\, _2F_1\left(\frac{\beta  (\nu -1)+1}{\beta  \nu },\frac{\mu +\nu }{\nu };\frac{\frac{1}{\beta }-1}{\nu }+2;-\left(\frac{q}{r_+}\right)^{\nu}\right)\Bigg\}.
  \end{align}
    \end{widetext}
    Within specific limits, the expression of the related black hole mass includes some other determined black hole masses. The situation of $\beta=0$ assists in the recovery of the specified mass for such a kind of AdS black hole in GR \cite{ElMoumni}, which when combined with $q =0$ produces the Schwarzschild-AdS black hole mass \cite{Dolan}.
    
The following step simply leads to finding the Hawking temperature. Taking into account the known surface gravity as 
\begin{equation}\label{r30}
    \kappa=\left(-\frac{1}{2}\nabla_\mu\xi_\nu\nabla^\mu\xi^\nu\right)^{1/2}=\frac{1}{2}f'\left(r_+\right),
\end{equation}
with $\xi^\mu=\partial/\partial t$  is a Killing vector. Thus, the formula $T_+=\kappa/2\pi$ leads to finding the Hawking temperature in the following way:
\begin{widetext}
    \begin{equation}\label{r31}
        T_+=\frac{1}{4\pi\, r_+}\Bigg(\frac{\Lambda  r_+^3}{4 \beta -1}+r_+-2r_+^3\left(\frac{q}{r_+}\right)^{\nu +3}\frac{\mu}{\alpha  \beta (\nu -1)+\alpha } \, _2F_1\left(\frac{\beta  (\nu -1)+1}{\beta  \nu
   },\frac{\mu +\nu }{\nu };\frac{\frac{1}{\beta }-1}{\nu }+2;-\left(\frac{q}{r_+}\right)^{\nu}\right)\Bigg).
    \end{equation}
    \end{widetext}
At some specific limits, the expression of the corresponding Hawking temperature yields alternative formulas. At the limit ($\beta \sim0$), the relevant Hawking temperature is confined to \cite{ElMoumni}, and when combined with $q =0$, the Hawking temperature is unmistakably that of the Schwarzschild-$AdS$ black hole \cite{Dolan}.
\begin{figure*}[htb]
      	\centering{
      	\includegraphics[scale=0.41]{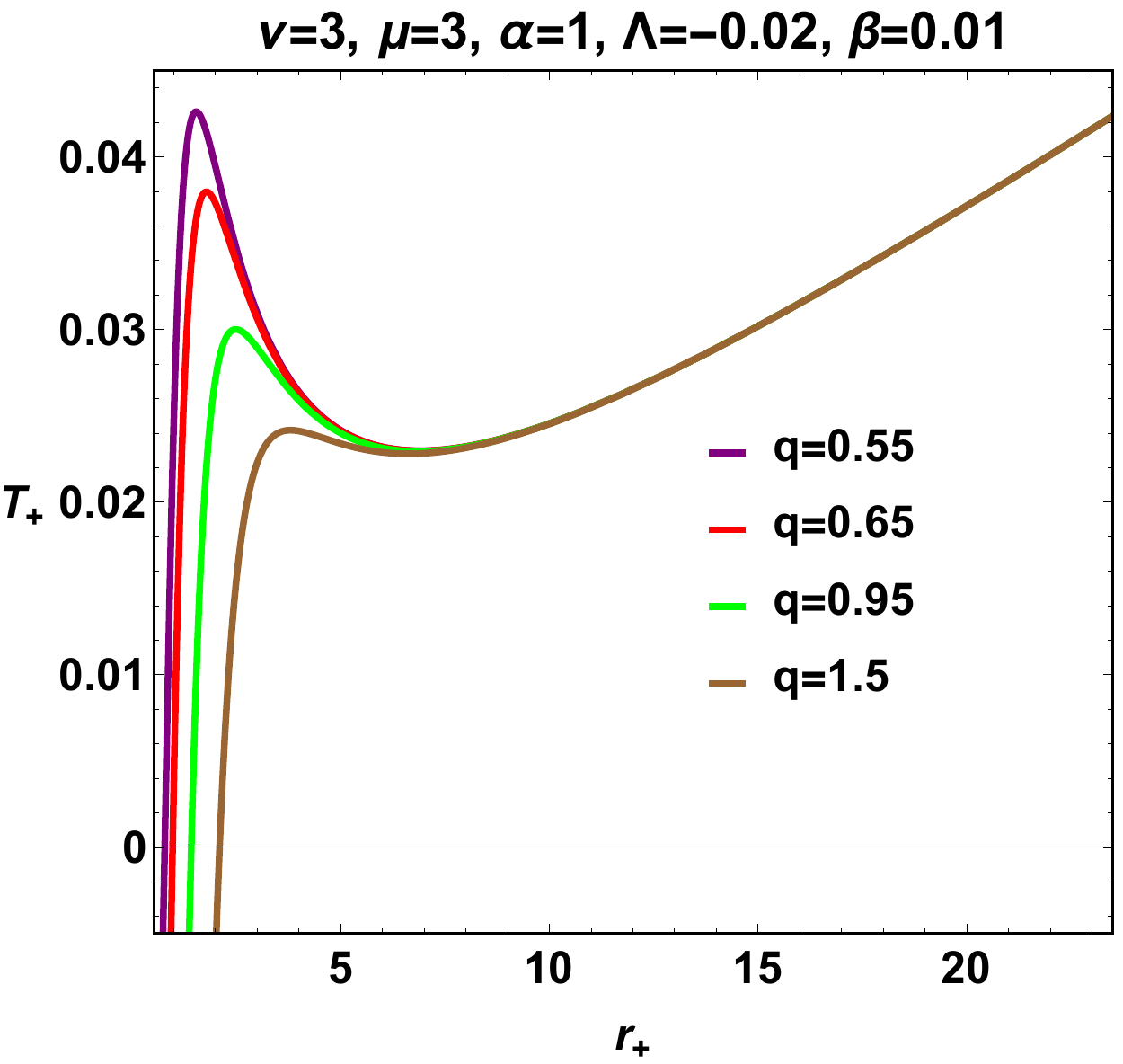} \hspace{0.15mm}
       \includegraphics[scale=0.42]{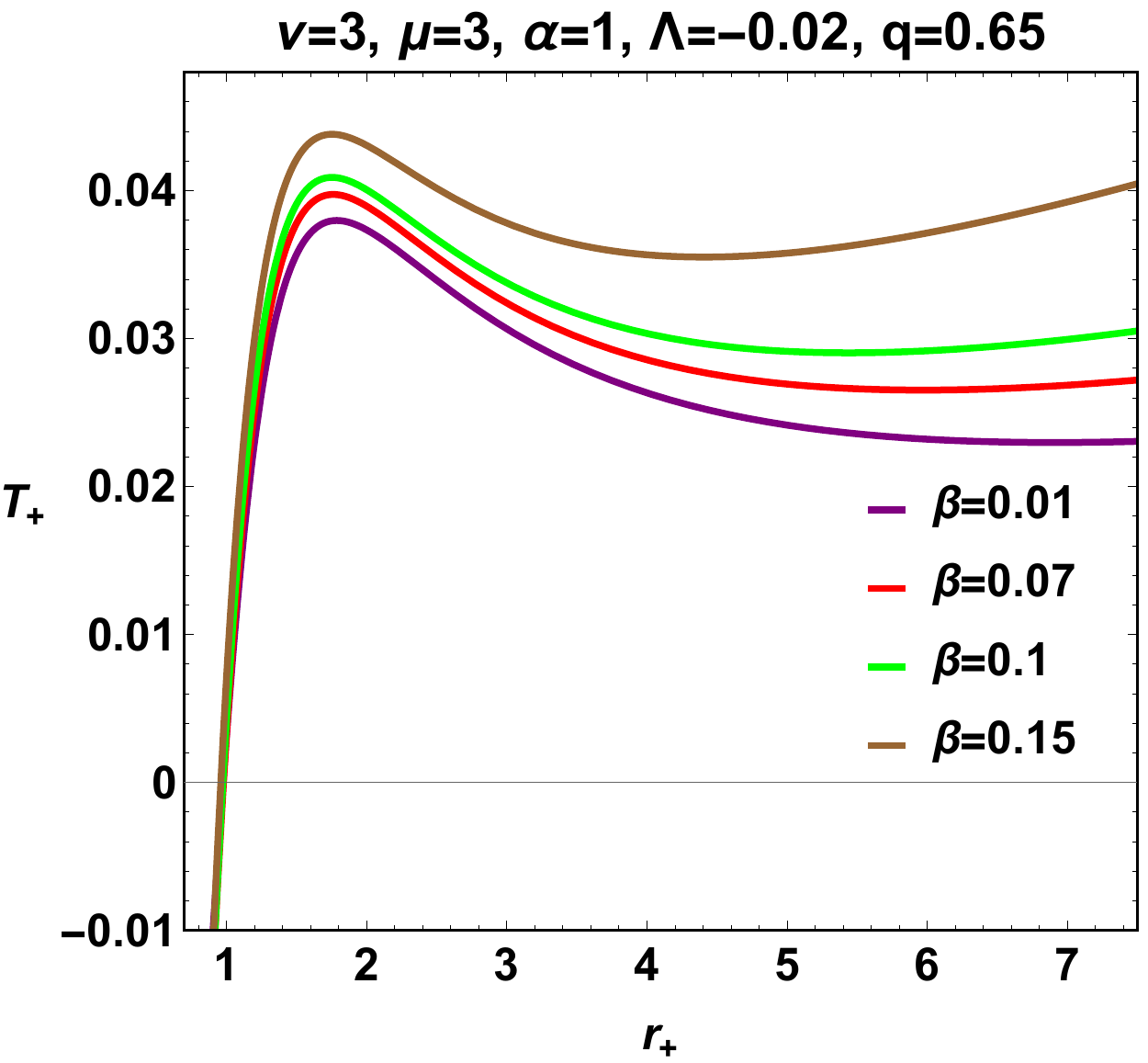}}
      	\caption{Variation of the Hawking temperature $T_+$ (\ref{r31}) with respect to $r_+$ for various values of $(q, \beta)$ and for fixed values of ($\nu, \mu, \alpha, \Lambda, \beta$).}
      	\label{fig2}
      \end{figure*}
To see how the Hawking temperature behaves when the parameter space is a set of fixed values, we plot this behavior in Fig. \ref{fig2}. We always chose to represent the variation in terms of the parameter pair $\left(q,\beta\right)$. Remarkably, the Hawking temperature rises to a maximum at $T^{max}_+$ for a specific point on the $r_+$ space, which grows depending on the variation of the parameter $\beta$. Roughly speaking, once the $\beta$-variation grows, the $T_+^{max}$ grows, in contrast to the $q$-variation, which is disproportional to the growth of the $T_+^{max}$. 
 { Moreover, further scrutiny reveals that the behavior of the Hawking temperature for essentially any parameter at small $r_+$ is unphysically due to the negativity dubbed $T+ < 0$. Accordingly, the presence of negative regions in the temperature diagram simply has no physical justification\cite{Feng:2015jlj,Feng:2017gms}, and according to many works\cite{Soroushfar:2021mis}, this region can be removed by introducing the cut-off length in such a way that it does not conflict with quantum gravity. }

It is advantageous to analyze and conduct a fascinating study for the thermodynamics aspect, keeping in mind the first law of thermodynamics in such a way that
\begin{equation}\label{r32}
    \mathrm{d}M=T_+ \mathrm{d}S+\sum_i \mu_i \mathrm{d}\mathcal{Q}_i
\end{equation}
where $\mu_i$ are the chemical potentials corresponding to the conserved charges $\mathcal{Q}_i$ \cite{Cai}.

Implementing the first law with the given equations of mass and temperature produces an equation for the corresponding entropy. Entropy is defined clearly by
\begin{equation}\label{r33}
    S=\frac{1}{T_+}\int\frac{\partial M_+}{\partial r_+}\mathrm{d}r_+=\pi r_+^2,
\end{equation}
where we notice that the related entropy satisfies the area-law entropy. 

To inspect the stability of the black hole system, certain tools are emerging for this study. Predicting the local stability of our black hole solution as a first task is only possible based on the sign of the heat capacity. This function is given by
\begin{equation}\label{r34}
C_+=\frac{\partial M}{\partial T_+}=\bigg(\frac{\partial M}{\partial r_+}\bigg)\bigg(\frac{\partial r_+}{\partial T_+}\bigg).
\end{equation}
Likewise, the function is expressed in terms of the parameter space as
 \begin{equation}\label{r35}
            C_+=2\pi\,\beta\left(\left(\frac{q}{r_+}\right)^{\nu }+1\right)^{\frac{\mu +\nu }{\nu }}\frac{C_1}{C_2}
        \end{equation}
        where
        \begin{widetext}
        \begin{align}\label{r36}
        C_1&= \alpha  q r_+ \left(\beta  (\nu -1)+1\right) \left(4 \beta +\Lambda  r_+^2-1\right)-2 \left(4 \beta -1\right) \mu  q^4 \left(\frac{q}{r_+}\right)^{\nu }
   \, _2F_1\left(\frac{\beta  (\nu -1)+1}{\beta  \nu },\frac{\mu +\nu }{\nu };\frac{\frac{1}{\beta }-1}{\nu }+2;-\left(\frac{q}{r_+}\right)^{\nu }\right)\nonumber\\
   C_2&=\Bigg(\beta  \bigg(\nu -1\bigg)+1\Bigg) \left(2\mu\,  q \bigg(4 \beta -1\bigg) \, r_+\, \left(\frac{q}{r_+}\right)^{\nu +2}-\alpha  \beta  \bigg(4 \beta -\Lambda  r_+^2-1\bigg) \left(\left(\frac{q}{r_+}\right)^{\nu }+1\right)^{\frac{\mu +\nu
   }{\nu }}\right)\Bigg(\frac{q}{r_+}\Bigg)\nonumber\\
   &+2\mu  q r_+\, \Bigg(\beta  \bigg(12 \beta -7\bigg)+1\Bigg)  \left(\left(\frac{q}{r_+}\right)^{\nu }+1\right) \left(\frac{q}{r_+}\right)^{\nu +3} \, _2F_1\left(1,\frac{\beta  (-\mu +\nu -1)+1}{\beta  \nu };\frac{\frac{1}{\beta
   }-1}{\nu }+2;-\left(\frac{q}{r_+}\right)^{\nu }\right).\nonumber
        \end{align}
        \end{widetext}
Within certain limits, as previously stated in the comments on mentioning thermodynamic variables such as mass and temperature, the corresponding heat capacity reduces. In both cases, $\beta=0$ and $q=0$, the heat capacity recovers that of \cite{Dolan}.

\begin{figure*}[tbh!]
      	\centering{
       \includegraphics[scale=0.254]{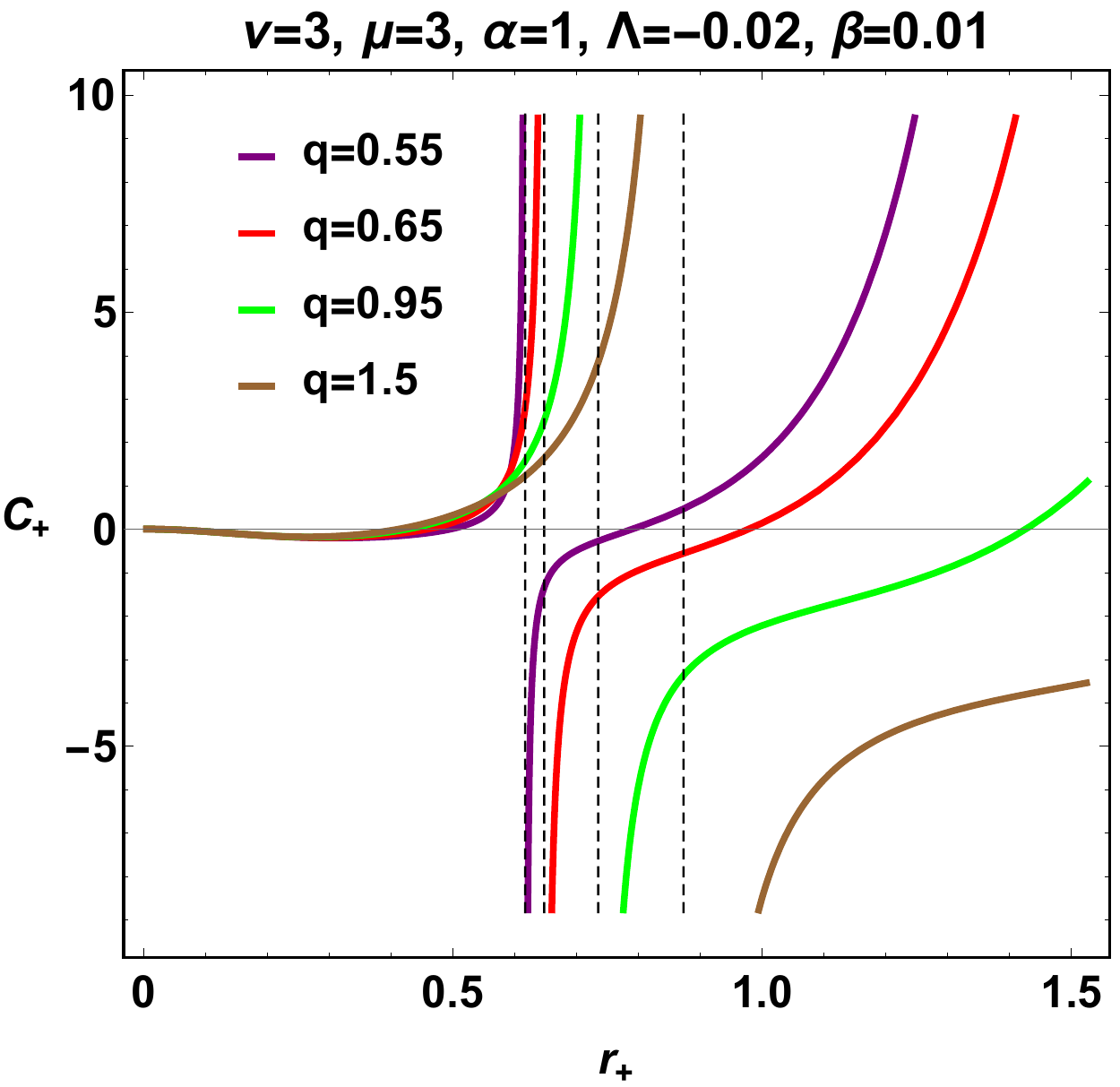} \hspace{2mm}
      	\includegraphics[scale=0.275]{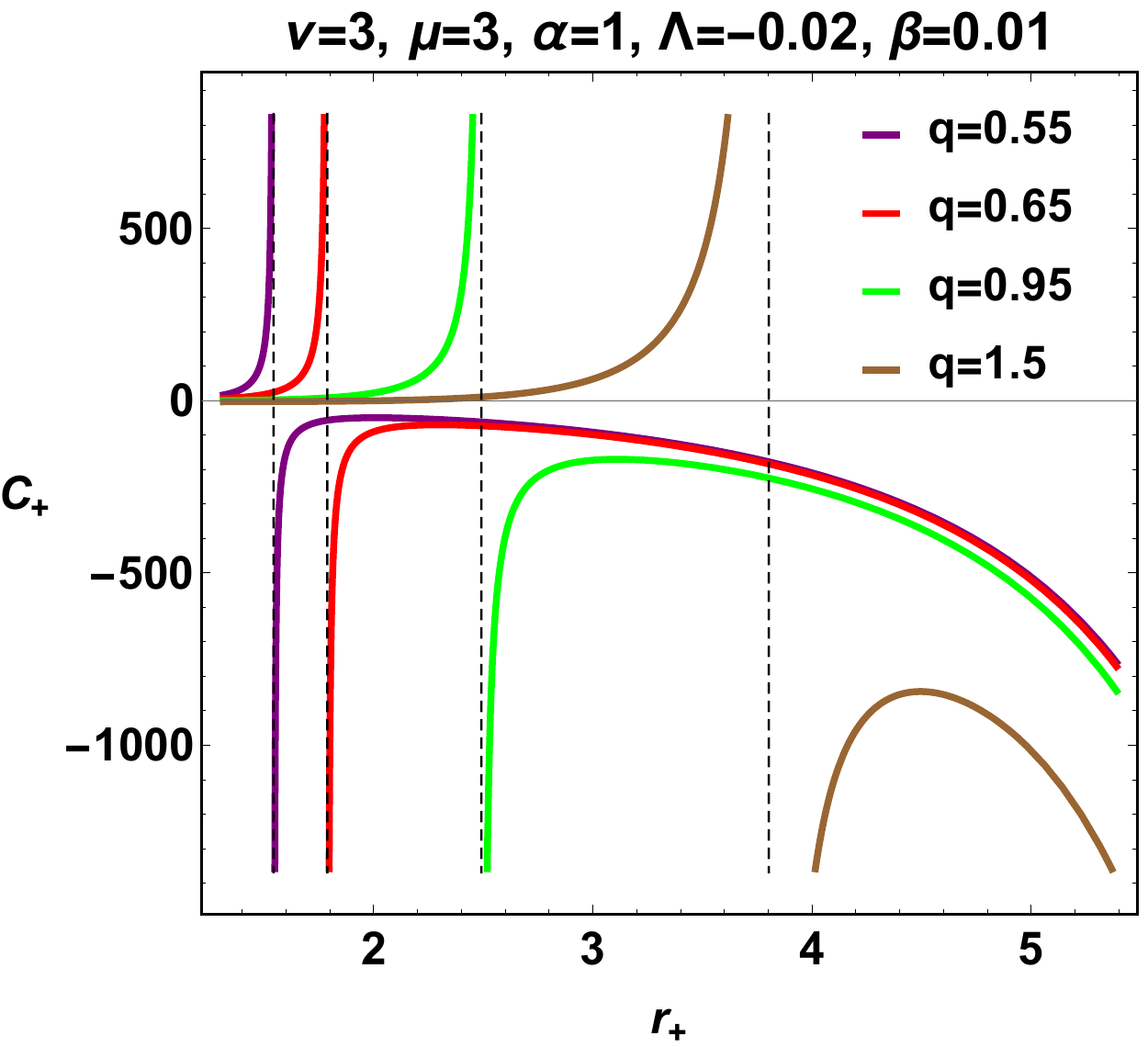} \hspace{2mm}
       \includegraphics[scale=0.283]{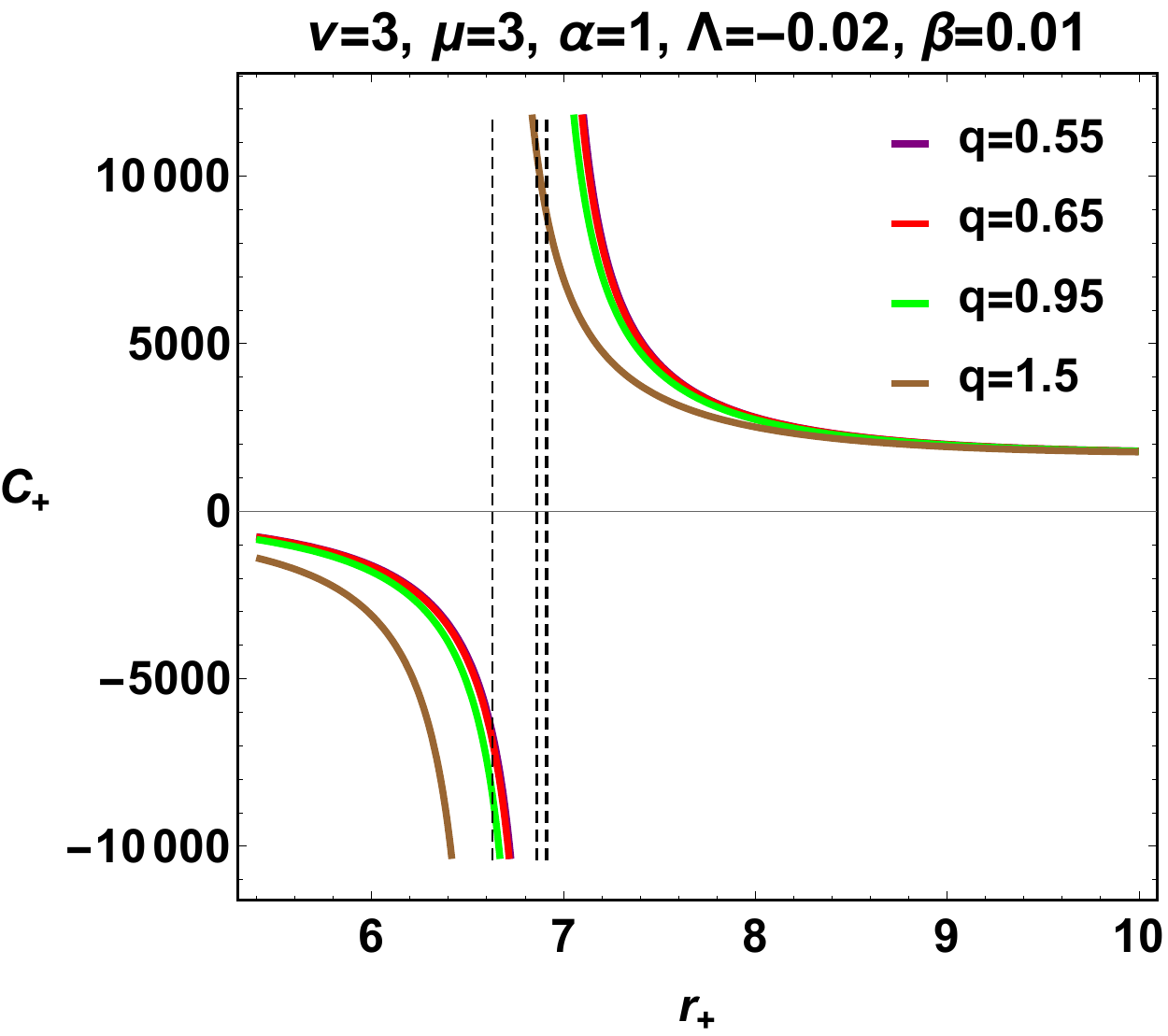}
      }
       \centering{ 
       \includegraphics[scale=0.254]{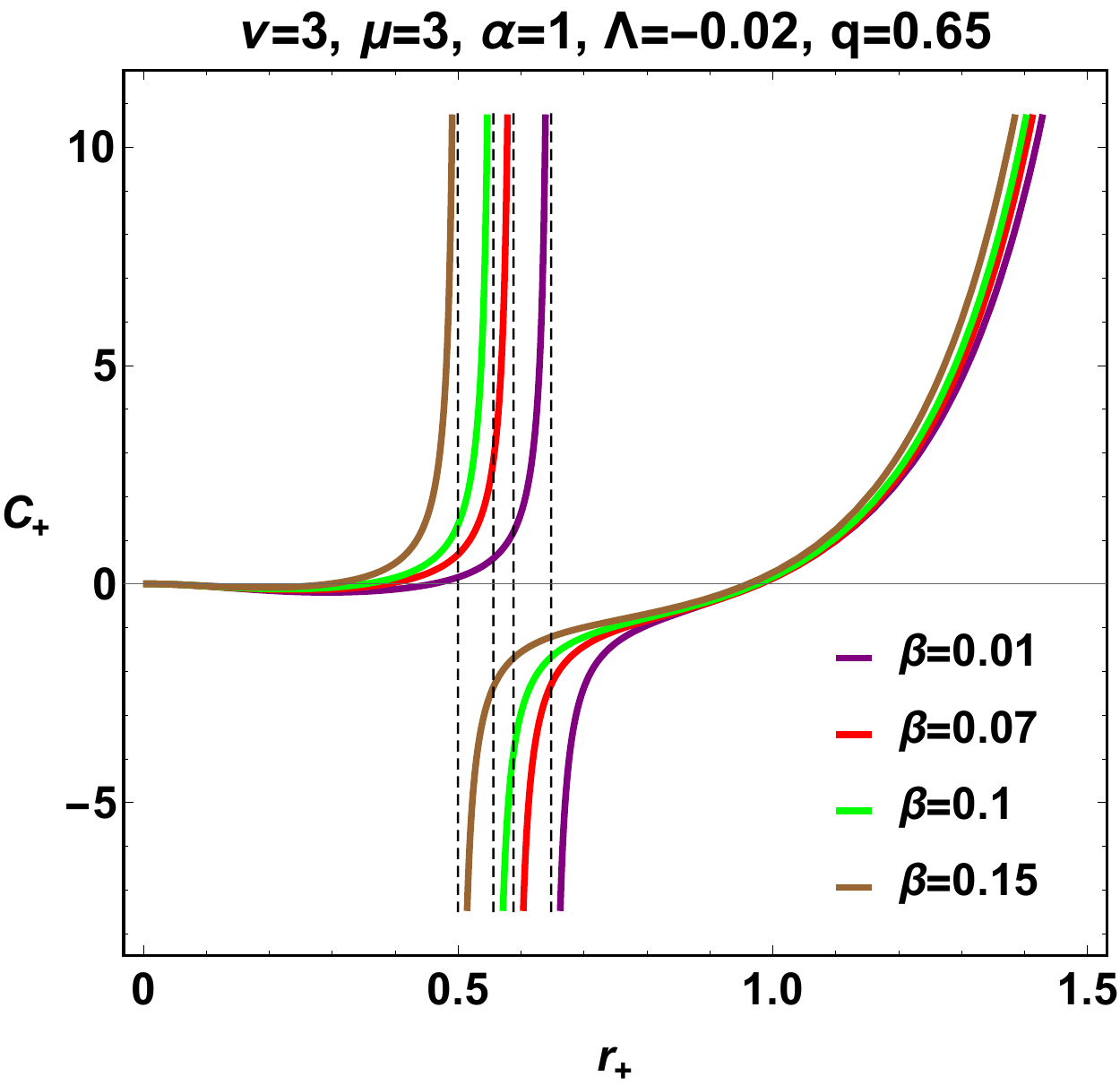} \hspace{2mm}
      	\includegraphics[scale=0.295]{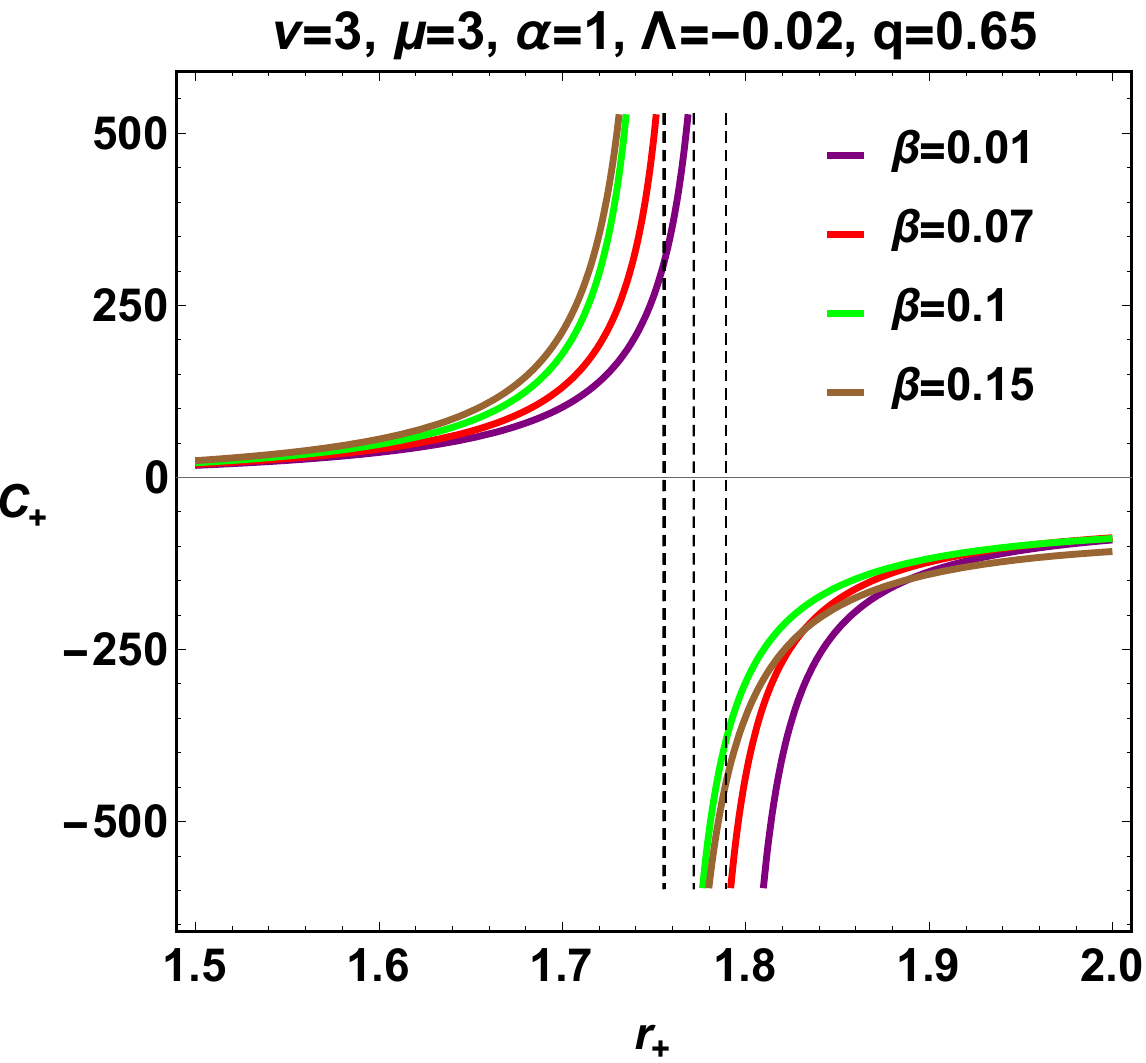} \hspace{2mm}
       \includegraphics[scale=0.31]{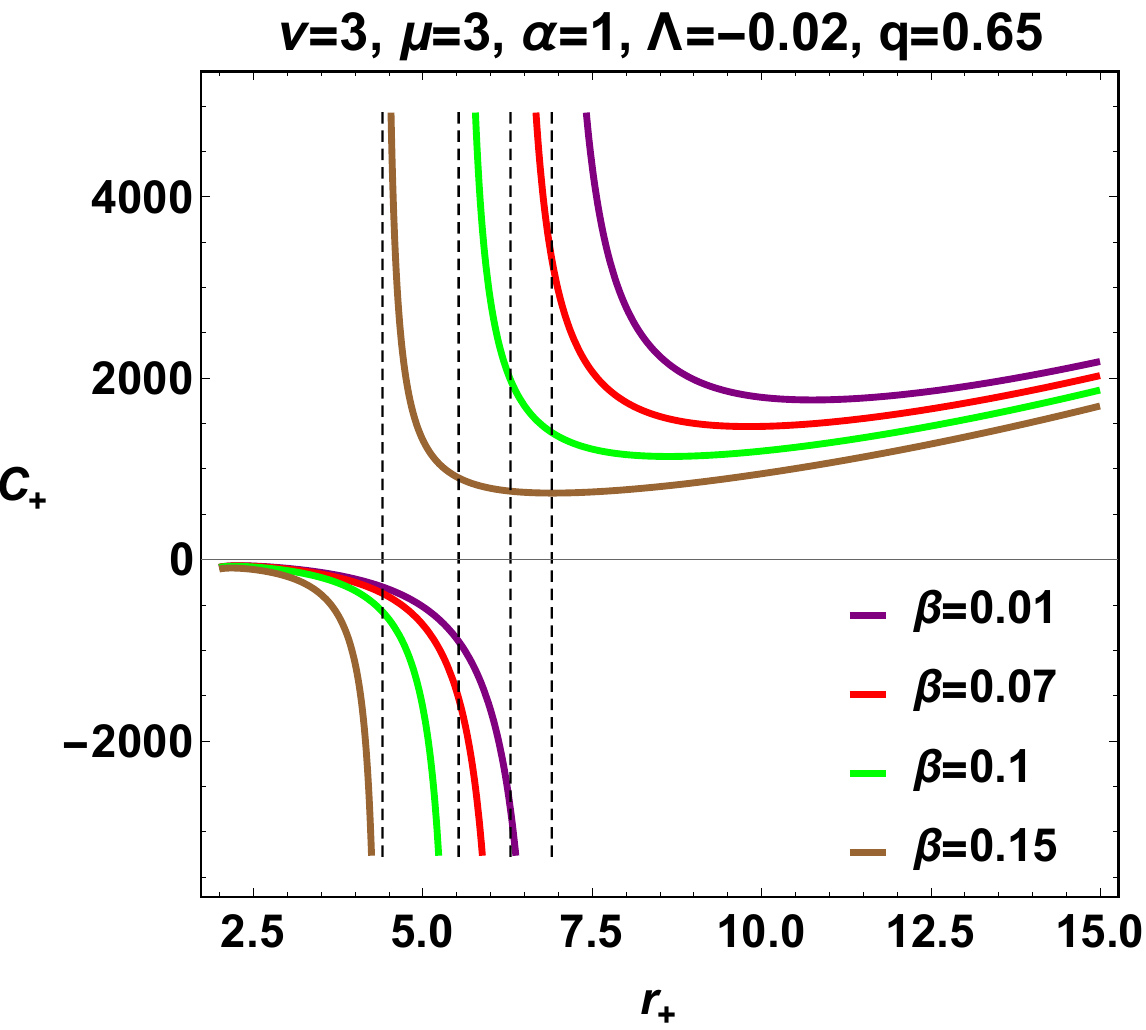}
       }
      	\caption{Variation of the Heat capacity $C_+$ (\ref{r34}) with respect to $r_+$ for various values of $(q, \beta)$ and for fixed values of ($\nu, \mu, \alpha, \Lambda, \beta$).}
      	\label{fig3}
      \end{figure*}

{ To better analyze the behavior of heat capacity, Fig. \ref{fig3} presents an appropriate analysis. It can be seen that the variation of parameters $q$ and $\beta$ generates a set of regions concerning the variation of heat capacity as a function of horizon radius with a specific sign. Thus, the three sub-figures for each fixed parameter space imply overall five regions in which three physical phase transition points (divergent points) and one physical limitation point (the root of heat capacity) are present. More specifically, the situation results in two types of sign changes of the heat capacity: either in a discontinuous way where the divergent point is concerned or by a continuous change generated by the physical limitation point. Essentially, the black hole system remains thermally locally stable since the sign of the heat capacity is positive.}  A close comparison between the pairs ($q$,$\beta$) shows that they have opposite behaviors on the heat capacity function.

The second task has the objective of predicting the global stability of our black hole solution. As a result, Gibbs free energy is the sole technique that involves specific global stability.
  { The appropriate Gibbs-free energy expression in normal phase space is given for this purpose as}
\begin{equation}\label{r37}
    F_+=M-T_+ S_+,
\end{equation}
or more precisely, in terms of the parameter space, the function $F_+$ is re-expressed as
    \begin{widetext}
    \begin{align}\label{r38}
        F_+&=\frac{1}{4} r_+ \Bigg\{1+2q\,r_+ \left(\frac{q}{r_+}\right)^{\nu +2} \frac{ (3 \beta -1) \mu   }{\alpha  (\beta -1) (\beta  (\nu -1)+1)} \, _2F_1\left(\frac{\beta  (\nu -1)+1}{\beta  \nu },\frac{\mu +\nu }{\nu };\frac{\frac{1}{\beta }-1}{\nu }+2;-\left(\frac{q}{r_+}\right)^{\nu}\right)+\frac{\Lambda\,  r_+^2}{3-12 \beta}\nonumber\\
   &+\frac{4q^3 }{\alpha  r_+ (1-\beta  r_+)}\left(1-\left(\left(\frac{q}{r_+}\right)^{\nu }+1\right)^{-\frac{\mu }{\nu }}\right)\Bigg\}.
   \end{align}
    \end{widetext}
In this case, the Gibbs free energy includes other expressions across the reducing parameter space. The cases ($\beta=0,q=0)$, in particular, lead to the knowing expressions. 
\begin{figure*}[htb]
      	\centering{
      	\includegraphics[scale=0.4]{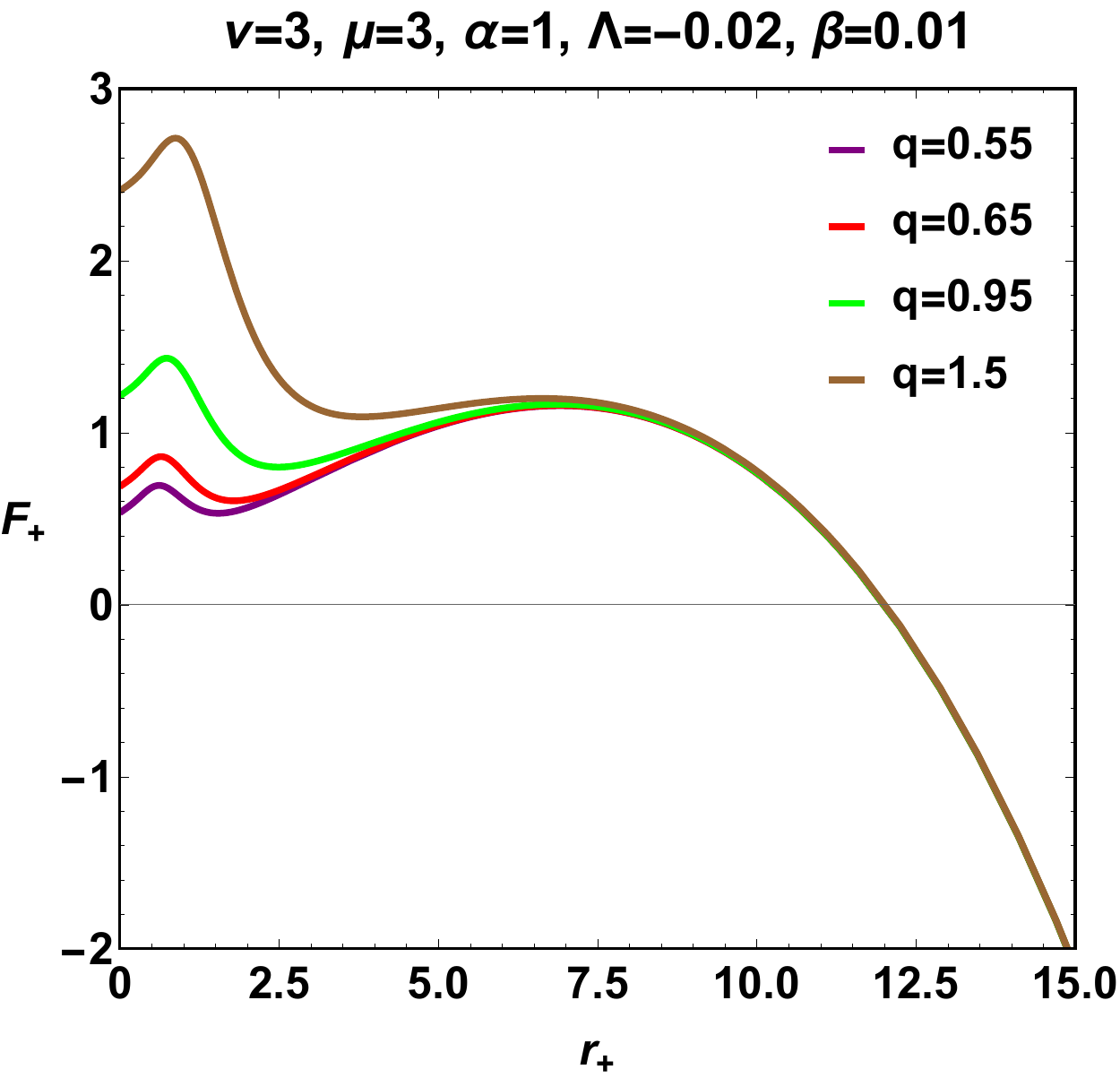} \hspace{5mm}
       \includegraphics[scale=0.4]{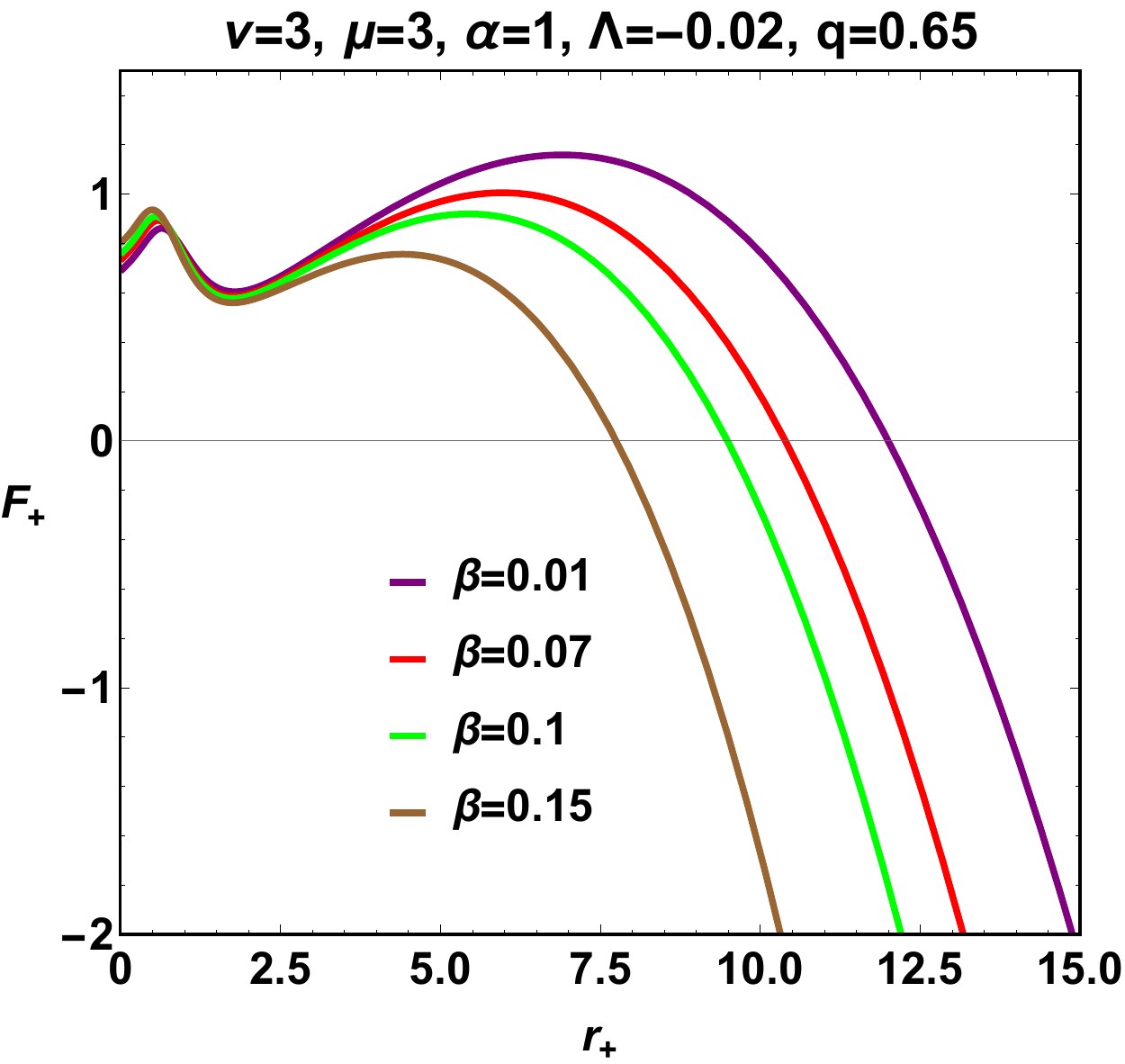}}
      	\caption{Variation of the Gibbs free energy $F_+$ (\ref{r38}) with respect to $r_+$ for various values of $(q, \beta)$ and for fixed values of ($\nu, \mu, \alpha, \Lambda, \beta$).}
      	\label{fig4}
      \end{figure*}
      { The usefulness of Gibbs-free energy demonstrates whether the black hole system is globally stable or unstable. For that reason, Fig. \ref{fig4} shows the variation of the Gibbs-free energy against the horizon radius, which claims that the black hole system is unstable at small $r_+$ $\left(F_+<0\right)$, while the positive sign of the Gibbs-free energy $\left(F_+>0\right)$ dominates, hence the black hole system remains globally stable}. The presence of such a minimum o maximum $\left(r_{min,max}\right)$, indicating the extreme point of the Hawking temperature on the other side, is a key feature in this figure. 
\section{Extended phase space}
The treatment of the $P$-$V$ critical behavior for $AdS$ generalized charged black holes in the extended phase space is the main objective of this section. Taking care of the $P$-$V$ criticality study is the best way to discover the critical process of the phase transition. The starting point for having an appropriate study of black hole chemistry is considering the following:
\begin{equation}\label{r49}
P=-\frac{\Lambda}{8\pi}
\end{equation}

To perform the intended phase transition, the first law of thermodynamics must be in the extended version. Thus, the first law is appropriately given as
\begin{equation}\label{r40}
\mathrm{d}H=\mathrm{d}M=T\mathrm{d}S+V\mathrm{d}P+\Phi_m \mathrm{d}Q_m+\Pi \mathrm{d}\alpha
\end{equation}
where $\Pi$ intervenes as a novel quantity conjugate to $\alpha$. Explicitly, it is defined by
\begin{equation}
    \Pi=\frac{1}{4}\int_{r_0}^\infty \mathrm{d}r\,\sqrt{-g}\,\frac{\partial\mathcal{L}}{\partial\alpha}.
\end{equation}
To construct the thermodynamic phase space framework, in particular, the parameter function $f(r_+, M,p,q,\alpha)$ must still vanish under any transformation of the parameter. Further remarks on this subject make similar arguments to consider the constraints $f(r_+, M,p,q,\alpha)= 0$ and $\delta f(r_+, M,p,q,\alpha)= 0$ on the evolution along the space of parameters. Nevertheless, an alternative way is considered: taking the mass parameter, $M$, as a function of the parameters $M(r_+,\ell, Q_m,\alpha)$ too.

The thermodynamic parameters are $S$, $p$, $Q_m$ and $\alpha$. It is then convenient to redefine $M=M(S,p,Q_m,\alpha)$ for the possibility of explicitly obtaining
\begin{widetext}
\begin{equation}\label{r41}
\mathrm{d}M=\bigg(\frac{\partial M}{\partial S}\bigg)_{P,Q_m,\alpha}\mathrm{d}S+\bigg(\frac{\partial M}{\partial P}\bigg)_{S,Q_m,\alpha}\mathrm{d}P+\bigg(\frac{\partial M}{\partial Q_m}\bigg)_{S,P,\alpha}\mathrm{d}Q_m+\bigg(\frac{\partial M}{\partial \alpha}\bigg)_{S,P,Q_m}\mathrm{d}\alpha.
\end{equation}
\end{widetext}
This is similar to such a differential 1-form in the space of parameters. Clearly, all of the components are nothing more than thermodynamic quantities, like temperature, thermodynamic volume, electric potential, and the parameter of the nonlinear electromagnetic field given respectively in the form
\begin{align}\label{r42}
T&=\bigg(\frac{\partial M}{\partial S}\bigg)_{P,Q_m,\alpha} \\
V&=\bigg(\frac{\partial M}{\partial P}\bigg)_{S,Q_m,\alpha}\\
\phi_m&=\bigg(\frac{\partial M}{\partial Q_m}\bigg)_{S,P,\alpha}\\
\Pi&=\bigg(\frac{\partial M}{\partial \alpha}\bigg)_{S,P,Q_m}.
\end{align}

In an alternative way, the same results can be expressed in accordance with the variation along the space of the parameters of the condition described by $f(r_+, M, P, Q_m,\alpha)$,
\begin{widetext}
\begin{equation}\label{r43}
\mathrm{d}f(r_+,M,P,Q_m,\alpha)=0=\frac{\partial f}{\partial r_+}\mathrm{d}r_++\frac{\partial f}{\partial M}\mathrm{d}M+\frac{\partial f}{\partial P}\mathrm{d}P+\frac{\partial f}{\partial Q_m}\mathrm{d}Q_m+\frac{\partial f}{\partial \alpha}\mathrm{d}\alpha
\end{equation}
\end{widetext}
reshape another term for $\mathrm{d}M$ giving as follows:
\begin{widetext}
\begin{align}\label{r44}
\mathrm{d}M&=\bigg(\frac{1}{4\pi}\frac{\partial f}{\partial r_+}\bigg)\bigg(-\frac{1}{4\pi}\frac{\partial f}{\partial M}\bigg)^{-1}\mathrm{d}r_++\bigg(-\frac{\partial f}{\partial M}\bigg)^{-1}\bigg(\frac{\partial f}{\partial P}\bigg)\mathrm{d}P+\bigg(-\frac{\partial f}{\partial M}\bigg)^{-1}\bigg(\frac{\partial f}{\partial Q_m}\bigg)\mathrm{d}Q_m\nonumber\\
&+\bigg(-\frac{\partial f}{\partial M}\bigg)^{-1}\bigg(\frac{\partial f}{\partial \alpha}\bigg)\mathrm{d}\alpha
\end{align}
\end{widetext}
which must conform with equation (\ref{r40}). Eq. (\ref{r44}) embraces the presence of temperature, which is geometrically defined as
\begin{equation}\label{r45}
T=\frac{1}{4\pi}\frac{\partial f}{\partial r_+},
\end{equation}
which is a well-known finding, providing
\begin{eqnarray}\label{r46}
\mathrm{d}S=\bigg(-\frac{1}{4\pi}\frac{\partial f}{\partial M}\bigg)^{-1}\,\mathrm{d}r_+.
\end{eqnarray}
It should be pointed out that this expression can also be derived using Wald's formalism; basically, $\delta S = \delta \int\frac{\partial L}{\partial R}$ as long as $\mathrm{d}f = 0$ is satisfied.

Furthermore, the thermodynamic volume, the electric potential, and the conjugate potential are defined by the following formula:
\begin{align}\label{r47}
V&=\bigg(\frac{\partial M}{\partial P}\bigg)_{S,Q_m,\alpha}=\bigg(-\frac{\partial f}{\partial M}\bigg)^{-1}\bigg(\frac{\partial f}{\partial P}\bigg)\\
\phi_m&=\bigg(\frac{\partial M}{\partial Q_m}\bigg)_{S,P,\alpha}=\bigg(-\frac{\partial f}{\partial M}\bigg)^{-1}\bigg(\frac{\partial f}{\partial Q_m}\bigg)\\
\Pi&=\bigg(\frac{\partial M}{\partial \alpha}\bigg)_{S,P,Q_m}=\bigg(-\frac{\partial f}{\partial M}\bigg)^{-1}\bigg(\frac{\partial f}{\partial \alpha}\bigg)
\end{align}
In the case of the black hole system, the enthalpy is defined by the total mass of the system. The thermodynamic volume can thus be computed as
\begin{equation}\label{r52}
V=\left(\frac{\partial M}{\partial P}\right)_{S,Q_m,\alpha}=\frac{4\pi r_+^3}{3}.
\end{equation}
According to Euler's theorem \cite{Altamirano}, with $M(S, P, Q_m,\alpha)$, the Smarr formula can be constructed for the charged source in Rastall gravity given by
\begin{equation}\label{r48}
M=2TS-2PV+\phi_m\,Q_m-2VP+2\Pi\,\alpha.
\end{equation}
in which the parameter $\alpha$ of the nonlinear electromagnetic field behaves as a thermodynamic variable.
\begin{figure*}[htb!]
      	\centering{
      	\includegraphics[scale=0.408]{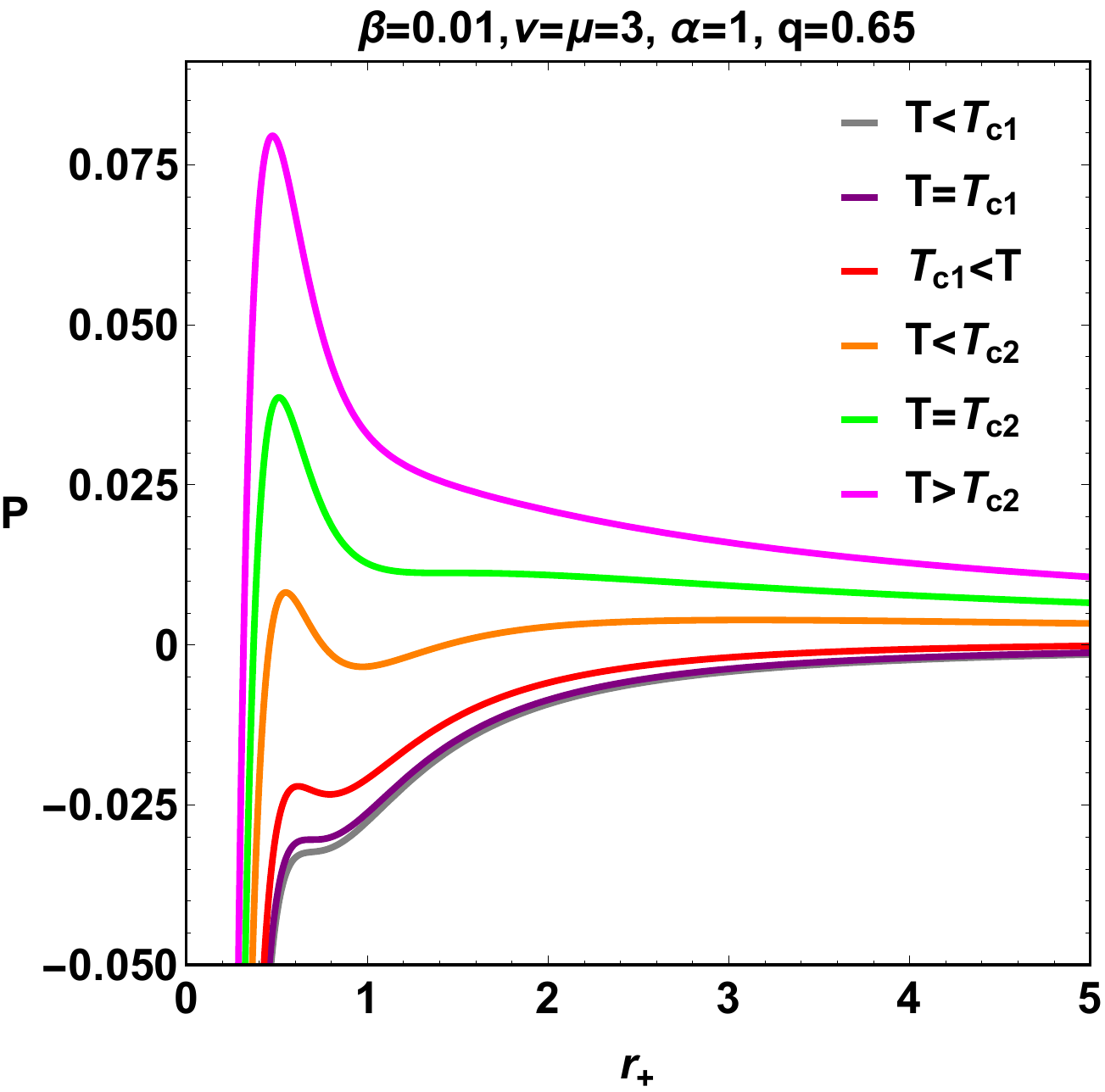} \hspace{0.3mm}\includegraphics[scale=0.4]{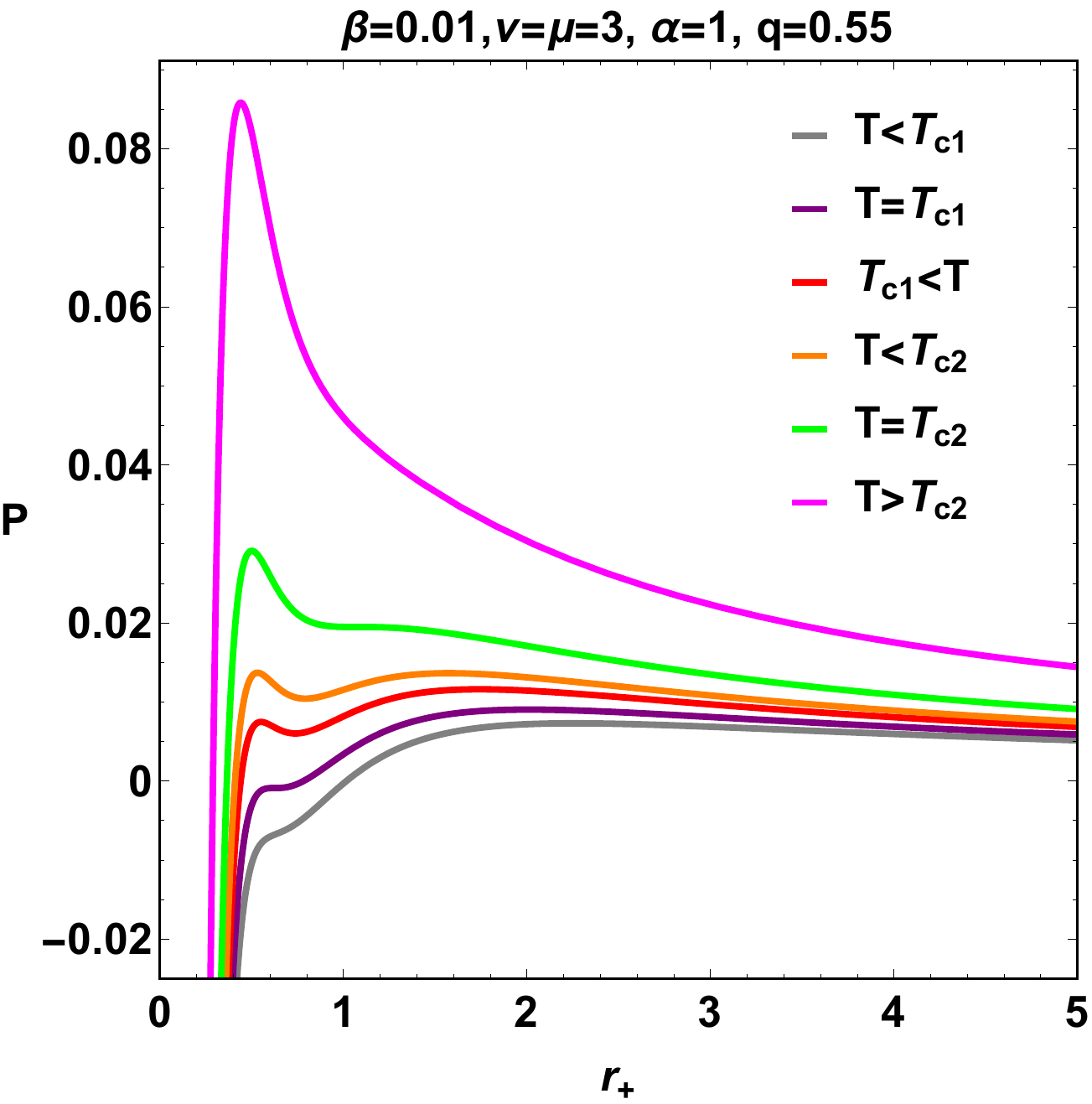}}
      	\caption{Isotherme $P$-$r_+$ of the black hole system for certain values of the parameter space}
      	\label{fig5}
      \end{figure*}

      Exploiting the Hawking temperature expression together with (\ref{r49}), we can obtain the corresponding equation of state for the black hole system in the following way:
\begin{widetext}
\begin{align}\label{r51}
P&=-\bigg(\frac{1-4\beta}{8\pi\, r_+^2}\bigg)+T\,\bigg(\frac{1-4\beta}{2r_+}\bigg)+\left(\frac{q}{r_+}\right)^{3+\nu}\frac{\left(1-4\beta\right)\mu}{4\pi\,\alpha\left(1+\beta(-1+\nu)\right)}\,_2F_1\left(\frac{\beta  (\nu -1)+1}{\beta  \nu },\frac{\mu +\nu }{\nu };\frac{\frac{1}{\beta }-1}{\nu }+2;-\left(\frac{q}{r_+}\right)^{\nu}\right).
\end{align}
\end{widetext}
Of course, the parameters $\beta$, $\alpha$, $\mu$ and $\nu$ potentially affect this equation. As intended, we can define the specific volume $v = 2r_+$, wherewith the pressure is cast in the standard form $P = \frac{1-4\beta}{v}\, T + \mathcal{O}(v)$. Moreover, since the thermodynamic volume $V \propto r^3_+$ the critical point can be determined by
\begin{equation}\label{r53}
\bigg(\frac{\partial P}{\partial r_+}\bigg)_T=0,\quad \bigg(\frac{\partial^2 P}{\partial r_+^2}\bigg)_T=0
\end{equation}
or alternatively,
\begin{equation}\label{r54}
\bigg(\frac{\partial T}{\partial r_+}\bigg)_P=0,\quad \bigg(\frac{\partial^2 T}{\partial r_+^2}\bigg)_P=0
\end{equation}
{ Using the constraints mentioned above permits the discovery of the critical configuration point $(T_c, P_c, r_c)$ which could be generated by considering the next equation with the unknown $r_c$ \eqref{inco}. So, the critical set can be defined as follows:
      \begin{widetext}
          \begin{align}
        T_c&=  -\frac{1}{2 \pi  r_c}\Biggl\{\frac{(4 \beta -1) \mu  2^{\frac{\nu -5}{8}} r_c^2 \left(\frac{\sqrt{\alpha }
   q^2}{r_c^4}\right){}^{\frac{\nu +3}{4}} }{\alpha  \beta  (\beta  (\nu -1)+1)}\, _2F_1\left(\frac{\beta  (\nu -1)+1}{\beta  \nu
   },\frac{\mu +\nu }{\nu };\frac{\frac{1}{\beta }-1}{\nu }+2;-2^{\nu /8} \left(\frac{q^2
   \sqrt{\alpha }}{r_c^4}\right){}^{\nu /4}\right)-1\nonumber\\
   &+\frac{\mu 
   2^{\frac{\nu -5}{8}} r_c^2 \left(\frac{\sqrt{\alpha } q^2}{r_c^4}\right){}^{\frac{\nu +3}{4}}
   \left(2^{\nu /8} \left(\frac{\sqrt{\alpha } q^2}{r_c^4}\right){}^{\nu
   /4}+1\right){}^{-\frac{\mu +\nu }{\nu }}}{\alpha  \beta }\Biggr\},
    \\
        P_c&= \frac{(1-4 \beta ) }{8 \pi 
   \alpha  r_c^2}\Biggl\{\alpha -\frac{(3 \beta -1) \mu  2^{\frac{\nu +3}{8}} r_c^2
   \left(\frac{\sqrt{\alpha } q^2}{r_c^4}\right){}^{\frac{\nu +3}{4}}}{\beta  (\beta  (\nu
   -1)+1)}\, _2F_1\left(\frac{\beta 
   (\nu -1)+1}{\beta  \nu },\frac{\mu +\nu }{\nu };\frac{\frac{1}{\beta }-1}{\nu }+2;-2^{\nu /8}
   \left(\frac{q^2 \sqrt{\alpha }}{r_c^4}\right){}^{\nu /4}\right)\nonumber\\
   &-\frac{\mu  2^{\frac{\nu +3}{8}} r_c^2 \left(\frac{\sqrt{\alpha }
   q^2}{r_c^4}\right){}^{\frac{\nu +3}{4}} \left(2^{\nu /8} \left(\frac{\sqrt{\alpha }
   q^2}{r_c^4}\right){}^{\nu /4}+1\right){}^{-\frac{\mu +\nu }{\nu }}}{\beta }\Biggr\}
      \end{align}
      and
\begin{align}\label{inco}
&2 \alpha  \beta ^3 \nu -\frac{\beta  (\beta  (12 \beta -7)+1) \mu 
   2^{\frac{\nu +3}{8}} \nu  r_c^2 \left(\frac{\sqrt{\alpha } q^2}{r_c^4}\right)^{\frac{\nu
   +3}{4}} }{\beta  (\nu -1)+1}\, _2F_1\left(\frac{\beta  (\nu -1)+1}{\beta  \nu },\frac{\mu +\nu }{\nu
   };\frac{\frac{1}{\beta }-1}{\nu }+2;-2^{\nu /8} \left(\frac{q^2 \sqrt{\alpha
   }}{r_c^4}\right)^{\nu /4}\right)\nonumber\\
   &+\beta  \mu  2^{\frac{\nu +3}{8}} \nu 
   r_c^2 \left(\frac{\sqrt{\alpha } q^2}{r_c^4}\right)^{\frac{\nu +3}{4}} \left(2^{\nu /8}
   \left(\frac{\sqrt{\alpha } q^2}{r_c^4}\right)^{\nu /4}+1\right)^{-\frac{\mu }{\nu }-2}
   \left(-\beta ( \nu +6 ) +2^{\nu /8} (\beta  (\mu -6)+1) \left(\frac{\sqrt{\alpha }
   q^2}{r_c^4}\right)^{\nu /4}+1\right)=0
\end{align}
\end{widetext}
Approaching numerically the critical behavior is achieved by solving Eq. \eqref{inco}. Thus, Tab. \ref{Tab1} collects, for various fixed parameter spaces, the numerically critical point $(T_{c_i}, P_{c_i}, r_{c_i})$. It is well observed that the associate critical set involves two physical critical points, and hence some complicated interpretations will appear compared to the standard case of the $P$--$V$ criticality.}

It is important to recall that in the limit $\beta\rightarrow0$ where the Rastall gravity is not present in the studied frame, the present state equation reduces to the one defined in general relativity expressed by
\begin{widetext}
\begin{equation}\label{r55}
P=-\frac{1}{8\pi r_+^2}+\frac{T}{2r_+}+\frac{\mu}{4\pi\alpha}\bigg(1+\left(\frac{q}{r_+}\right)^\nu\bigg)^{-\frac{\mu+\nu}{\nu}}\left(\frac{q}{r_+}\right)^{\nu+3}.
\end{equation}
\end{widetext}
Exploiting the above equation with condition 32, the critical point will be explicitly defined as
\begin{itemize}
    \item type $(I)$:
\end{itemize}
\begin{center}
$r_c=2.48516\,q$ $\quad$ $T_c=\frac{0.04313\,q}{\alpha}$ $\quad$ $P_c=\frac{0.007055}{\alpha}$
\end{center}
\begin{itemize}
    \item type $(II)$:
\end{itemize}
\begin{center}
\hspace{0.8cm}$r_c=3.1454\,q$ $\quad$ $T_c=\frac{0.0205236\,q}{\alpha}$ $\quad$ $P_c=\frac{0.00248808}{\alpha}$
\end{center}
\begin{itemize}
    \item Maxwellian type:
\end{itemize}
\begin{center}
    $r_c=5.449\,q$ $\quad$ $T_c=\frac{0.00311\,q}{\alpha}$ $\quad$ $P_c=\frac{0.0001903}{\alpha}$
\end{center}
where finding the root as $r_c$ leads to the definition of the critical triplet $(r_c, T_c, P_c)$.

\begin{figure*}[tbh!]
      	\centering{
       \includegraphics[scale=0.46]{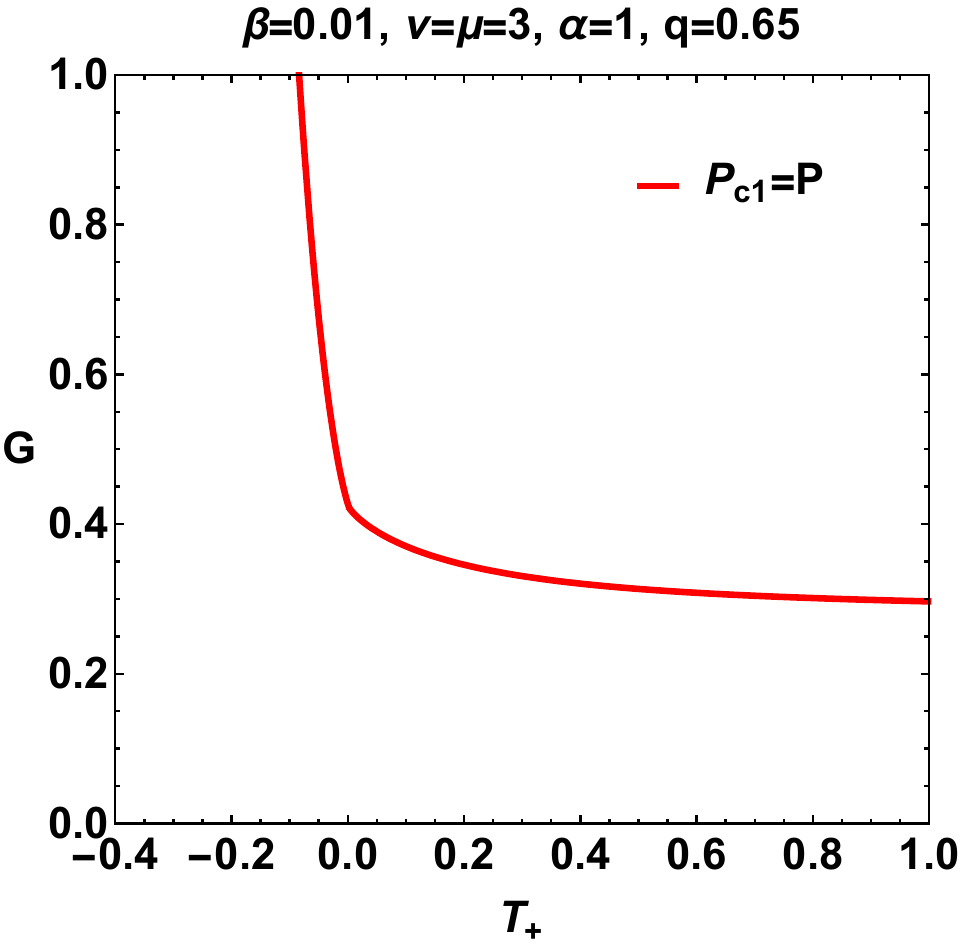} \hspace{2mm}
      	\includegraphics[scale=0.45]{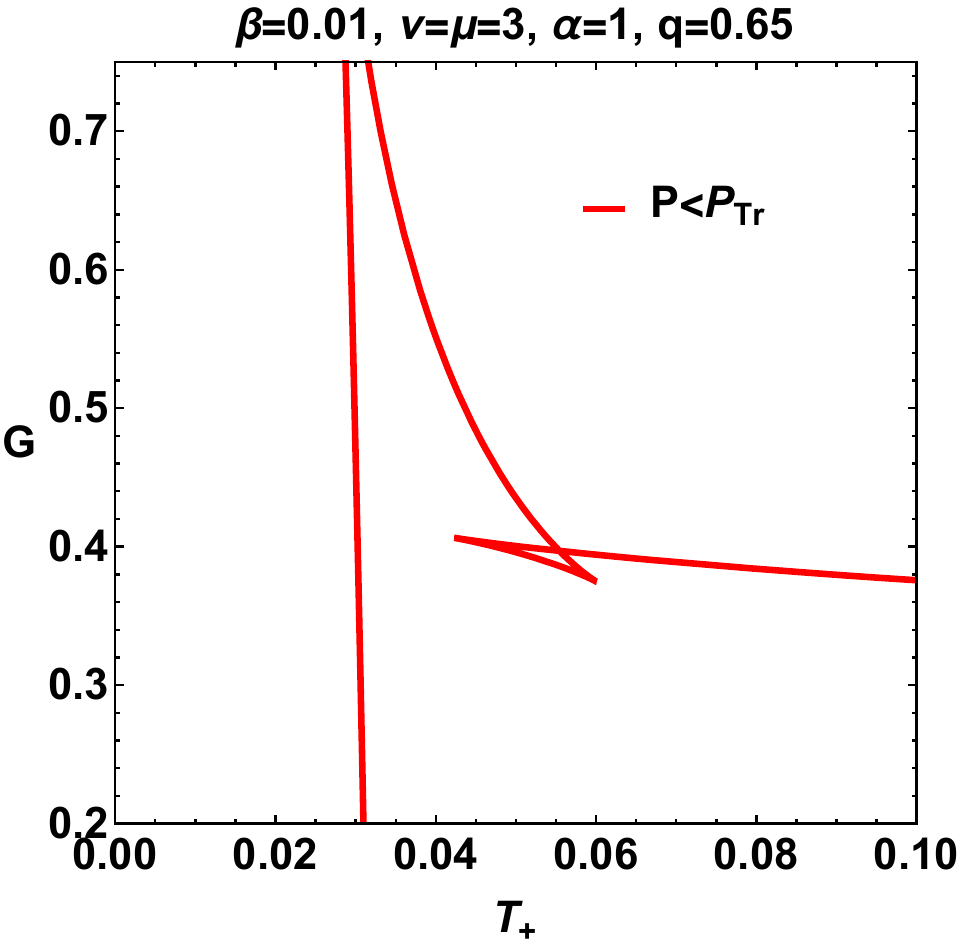} \hspace{2mm}
       \includegraphics[scale=0.45]{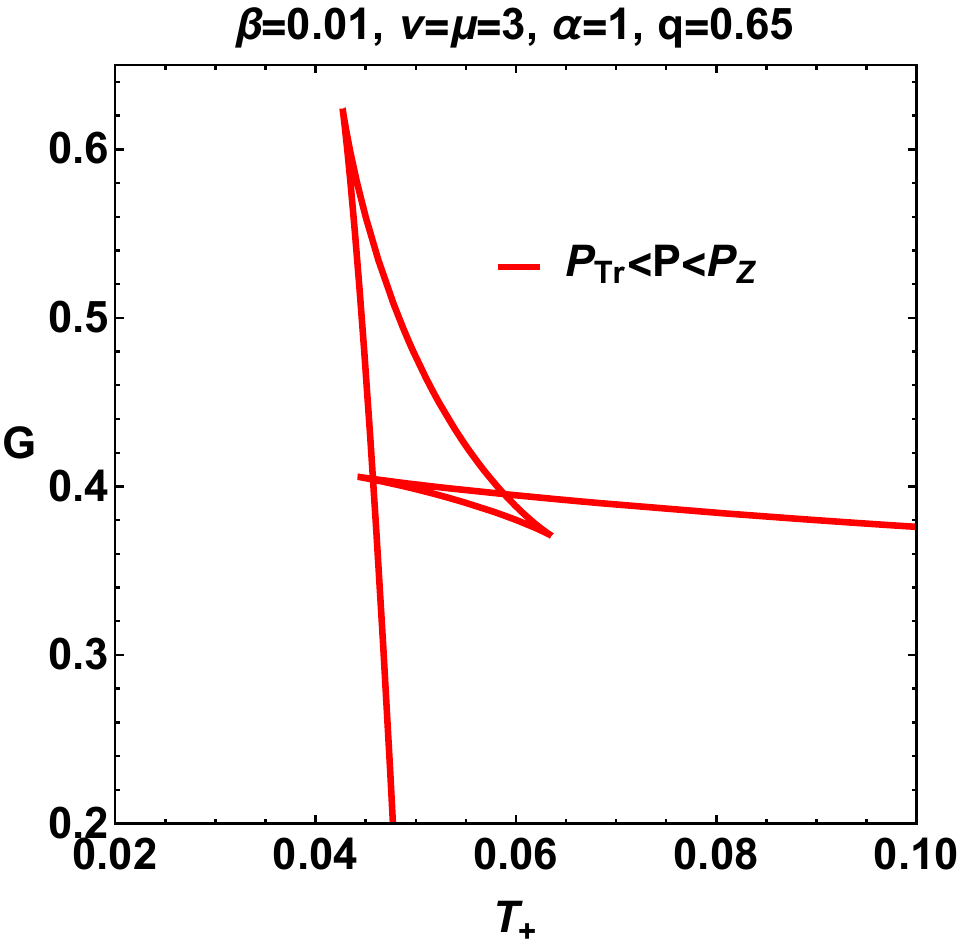}}
       \centering{
      \includegraphics[scale=0.45]{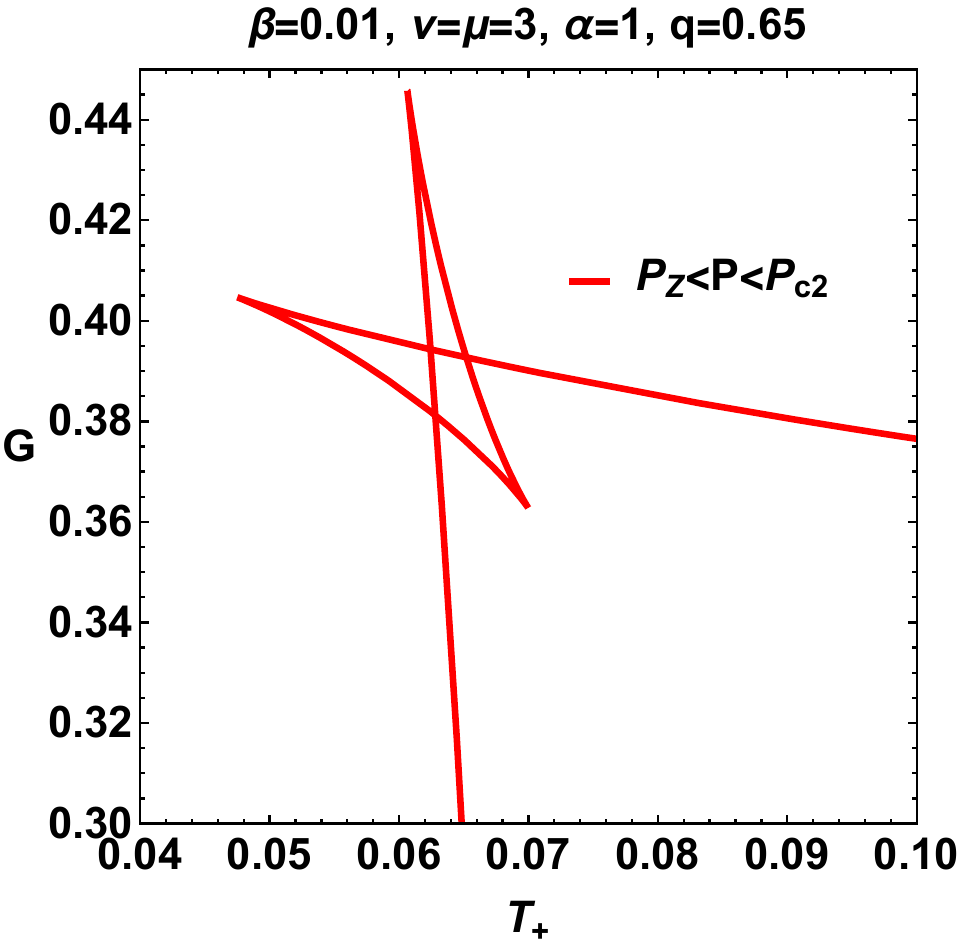} \hspace{5mm}
      \includegraphics[scale=0.45]{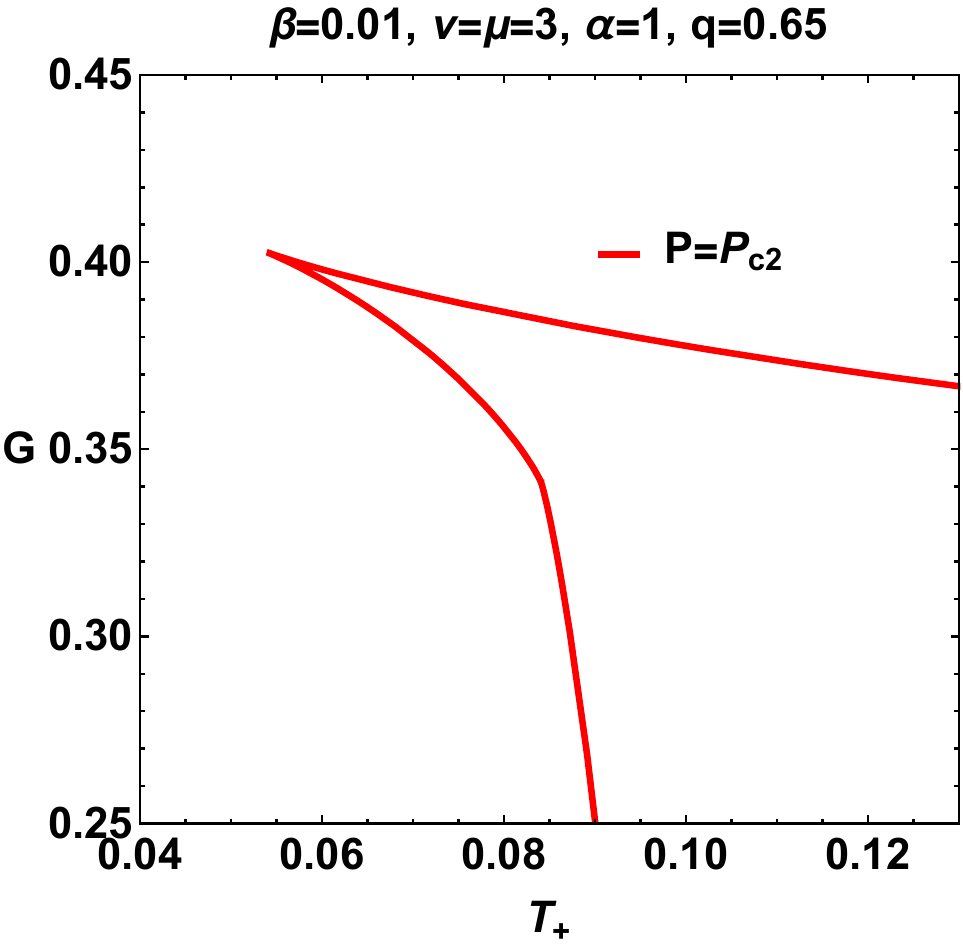} \hspace{5mm}
      \includegraphics[scale=0.45]{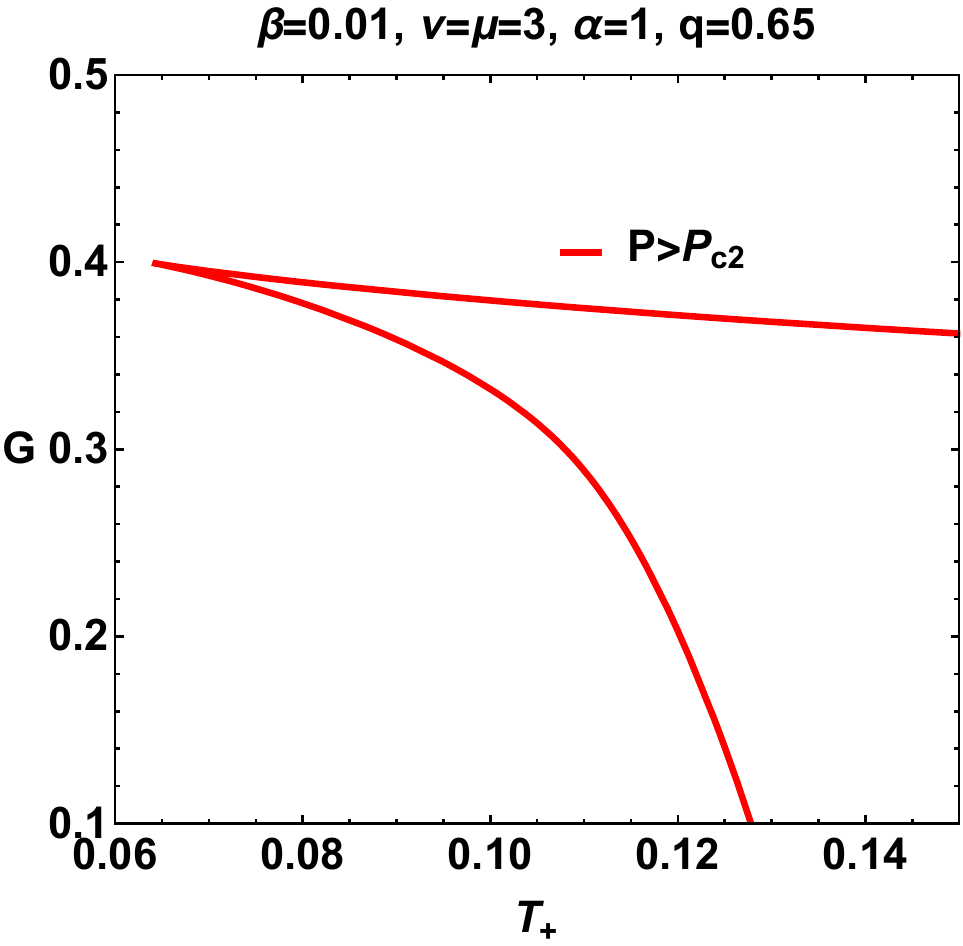}
       }
      	\caption{The $G$ - $T$ diagram of the black hole system for certain values of the parameter space.
Critical data: ($T_{Tr} = 0.0448$, $P_{Tr} = 0.0024087$) and ($T_Z = 0.0573$, $P_Z = 0.00425724$)}
      	\label{fig6}
      \end{figure*}

To evidence the $P$-$r_+$ criticality behavior, we plot the $P$-$r_+$ diagram in Fig. \ref{fig5} for various values of Hawking temperature $T$. Fig. \ref{fig5} shows that for certain values of the parameter space, there exist two critical points: one refers to negative (unphysical) pressure and one to positive pressure. The obtained curves on the $P$-$r_+$ diagram are moderately more complex than those of the standard van der Waals, reflecting the important behavior of Gibbs free energy. Thus, at a small value of $r_+$, the isotherms behave like van der Waals behavior, while at a large value of $r_+$, the isotherms turn and lead back to a region with negative pressures. The isothermal plots regarding this situation are perfectly similar to the $P$-$V$ diagram of Born-Infeld-AdS black holes \cite{Born}. A comparative study of the parameter charge shows that once the charge parameter increases, the maximum along the $P$-$r_+$ curves decreases.

The next step will be to perfectly disclose the corresponding phase transition of a system by means of its thermodynamic potential. In particular, Gibbs's free energy is a thermodynamic quantity computed from the Euclidean action with an appropriate boundary term. A fruitful feature resulting from the sign of Gibbs free energy enables an analysis of global stability. In the extended phase space, the thermodynamic potential is now Gibbs free energy $G=M-T\, S$. In practice, it is worth noting that any discontinuous behavior in first- or second-order derivatives of Gibbs energy results in a first- or second-order phase transition in the system. 
 { So, the appropriate Gibbs-free energy expression in extended phase space is given for this purpose as}
\begin{widetext}
\begin{align}\label{r56}
G&=G(P, T)=\frac{r_+}{4}\Bigg\{1+r_+^2\Bigg[ \frac{8\pi}{-3+12\beta}P+\frac{4}{\alpha(1-\beta)}\bigg(1-\left(1+\left(\frac{q}{r_+}\right)^\nu\right)^{-\frac{\mu}{\nu}}\bigg)\left(\frac{q}{r_+}\right)^3\Bigg]\nonumber\\
&+2\left(\frac{q}{r_+}\right)^{3+\nu}\frac{(-1+3\beta)\mu\, r_+^2}{\alpha(-1+\beta)\left(1+\beta(-1+\nu)\right)}\, _2F_1\left(\frac{\beta  (\nu -1)+1}{\beta  \nu },\frac{\mu +\nu }{\nu };\frac{\frac{1}{\beta }-1}{\nu }+2;-\left(\frac{q^4}{r_+^4}\right)^{\nu
   /4}\right)\Bigg\}.
\end{align}
\end{widetext}
The appropriate Gibbs behavior for Fig. \ref{fig5} with two critical points, is shown in Fig. \ref{fig2}. In a detailed discussion, the figure at the top left shows the behavior of the $G$-$T$ curve for $P_{c1}=-0.0803596<0$, which is a thermodynamically unstable branch with negative heat capacity $C_p<0$. In the $P<P_{Tr}$ range, a new branch of thermodynamically stable large black holes appears, in which two zero-order phase transitions do not minimize the Gibbs free energy. As a result, only one phase of large black holes is present. Furthermore, in the next range associated with $P_{Tr}<P<P_Z$, here, the minimum of Gibbs free energy is discontinuous. At that range, two phases of intermediate-sized black holes exist. These phases are joined by a jump in G or a zero-order phase transition. The underlying feature in this range is that the critical behavior admits an ordinary reentrant large/small/large black hole phase transition. Moreover, once $P=P_Z$, the reentrant phase transition phenomenon is no longer there spontaneously, but the first-order phase transition remains present. For $P_Z<P<P_{c2}=0.0028767$, a first-order phase transition takes place, similar to Van der Waals fluid. At $P=P_{c2}$, we have the second-order phase transition, and thus the second critical point is physical. Moreover, for $P>P_{c2}$, the system displays no phase transitions. Therefore, the critical point with negative pressure does not globally minimize the Gibbs energy, which is not a physical point, while the second critical point minimizes the Gibbs free energy.

\begin{table*}[ht!]
    \begin{tabular}{lcccc} 
    \hline\hline
        $q$  \,& $\beta$ \,& $r_{c1}/r_{c2}$ \,&  $T_{c1}/T_{c2}$ \,& $P_{c1}/P_{c2}$
       \\
       \hline
       0.55\,&\,0.01 \,&\, 0.90498/1.63727 \,&\, 0.0424235/0.0732472 \,&\, -0.00417202/0.00853251 \\
       0.55\,&\,0.03 \,&\, 0.87897/1.63013 \,&\, 0.0379238/0.0737712 \,&\, -0.00589315/0.00792973 \\
       0.65 \,&\, 0.01 \,&\, 0.952804/1.89952 \,&\, 0.0162237/0.0638822\,&\,-0.0115852/0.0064755 \\
       0.65\,&\,0.03 \,&\, 0.926721/1.88874 \,&\, 0.0104719/0.0643721 \,&\, -0.0130851/0.0060249\\
           \hline\hline
    \end{tabular}
     \caption{{ Numerical sets for critical physical quantities $(r_c, T_c, P_c)$ with $\nu =\mu =3$ and $\alpha= 1$.}}
      \label{Tab1}
    \end{table*}
    
   \begin{table*}[ht!]
    \begin{tabular}{lcccc} 
    \hline\hline
        $q$  \,& $\beta$ \,& $a$ \,&  $a_1(=-a_3)$ \,& $a_5$
       \\
       \hline
       0.55\,&\,0.01 \,&\, 1 \,&\, -5.393595/2.51672 \,&\, -17.2554/-1.35435 \\
       0.55\,&\,0.03 \,&\,1 \,&\, -3.22138/2.51107 \,&\, -12.4961/-1.38149 \\
       0.65 \,&\, 0.01 \,&\, 1 \,&\, -0.70548/2.49289\,&\,-7.10661/-1.46898 \\
       0.65\,&\,0.03 \,&\, 1 \,&\, -0.379941/2.48398 \,&\, -6.35601/-1.48885\\
           \hline\hline
    \end{tabular}
     \caption{{ Numerical sets for the constants $a_i$ with $\nu =\mu =3$ and $\alpha= 1$.}}
      \label{Tab2}
    \end{table*}
    
{ 
\subsection{Critical exponents}
Critical exponents neatly portray the behavior of physical quantities in the vicinity of the critical point. This is because critical exponents are independent of physical systems and can be considered as quasi-universal parameters. It may help to consider the following key notations:
\begin{equation}
    t=\frac{T}{T_c}-1,\quad\omega=\frac{V}{V_c}-1,\quad p=\frac{P}{P_c},
\end{equation}
wherein the critical thermodynamic volume $V_c$ is assigned to the critical horizon radius $r_c$ by $V_c=\frac{r_c^4}{2}$. In this way, the critical exponents are spelled out explicitly as follows:
\begin{align}
    C_V&\propto\lvert t\rvert^{-\alpha}\\ \eta&\propto\lvert t\rvert^{\lambda}\\
    \kappa_T&\propto\lvert t\rvert^{-\gamma}\\ \lvert P-P_c\rvert&\propto\lvert V-V_c\rvert^\delta.\label{71}
    \end{align}

    The exponent $\alpha$ specifies the behavior of specific heat at constant volume. A simple conclusion is that the entropy $S$ is independent of the Hawking temperature $T$, hence
    \begin{equation}
        C_V= T\left(\frac{\partial S}{\partial T}\right)_V=0,
    \end{equation}
so we can deduce that $\alpha = 0$.

   The exponent $\beta$ captures the behavior of the order parameter near the critical point. In this case, the equation of state near the critical point can be developed as 
   \begin{widetext}
        \begin{equation}\label{pr}
         p=a+a_1\, t+a_2\, \omega+a_3\,t\omega+a_4\,\omega^2+a_5\,\omega^3+O(t\omega^2,\omega^4),
     \end{equation}
   \end{widetext}
     where
     \begin{widetext}
          \begin{align}
         a& =\frac{\left(1-4 \beta \right) }{8 \pi  \alpha  P_c r_c^2}\left(\frac{\mu  2^{\frac{\nu +3}{8}} r_c^2  }{\beta  (\nu -1)+1}\left(\frac{\sqrt{\alpha } q^2}{r_c^4}\right)^{\frac{\nu
   +3}{4}}\,\mathcal{U}+\alpha  \left(4 \pi 
   r_c T_c-1\right)\right),\nonumber\\
  a_1 &=-a_3=\frac{ T_c}{2 P_c r_c}(1-4 \beta ) ,\nonumber\\
   a_2& =a_4 =0,\nonumber\\
   a_5& =\frac{\left(4 \beta -1\right)}{24 \pi  \alpha 
   \beta ^2 P_c}\Biggl\{ \Bigg(\frac{1}{\beta }\bigg(\mu  2^{\frac{\nu -5}{8}} \left(\frac{\sqrt{\alpha }
   q^2}{r_c^4}\right)^{\frac{\nu +3}{4}} \bigg(\frac{\beta ^3 (\nu +3) (\nu +4) (\nu +5) }{\beta 
   (\nu -1)+1}\mathcal{U}-\frac{1}{\beta  (2 \nu
   -1)+1}\nonumber\\
   &\times\bigg(\beta  2^{\nu /8} \bigg(\beta ^2 (\nu  (\nu +13)+60)-\beta  (\nu +14)+1\bigg)
   (\mu +\nu ) \left(\frac{\sqrt{\alpha } q^2}{r_c^4}\right)^{\nu /4} \mathcal{V}\bigg)\bigg)\bigg)+\mu  2^{\frac{1}{8} (2 \nu -5)} (\mu +\nu )\nonumber\\
   &\times \left(\frac{\sqrt{\alpha }
   q^2}{r_c^4}\right)^{\frac{\nu }{2}+\frac{3}{4}} \frac{\bigg(-3 \beta  \nu -13 \beta
   +2^{\nu /8} (\beta  (\mu -\nu -13)+1) \left(\frac{\sqrt{\alpha } q^2}{r_c^4}\right)^{\nu
   /4}+1\bigg)}{\left(2^{\nu /8} \left(\frac{\sqrt{\alpha}
   q^2}{r_c^4}\right)^{\nu /4}+1\right)^{\frac{\mu }{\nu }+3}} +\frac{12 \alpha  \beta ^2 \left(\pi  r_c T_c-1\right)}{r_c^2}\Bigg)\Biggr\},\nonumber
     \end{align}
     
     \end{widetext}
     with 
     \begin{widetext}
          \begin{align}
         \mathcal{U}&= \,_2F_1\left(\frac{\beta  (\nu -1)+1}{\beta  \nu },\frac{\mu +\nu }{\nu };\frac{\frac{1}{\beta }-1}{\nu
   }+2;-2^{\nu /8} \left(\frac{q^2 \sqrt{\alpha }}{r_c^4}\right)^{\nu /4}\right),\nonumber\\
\mathcal{V}& = \,_2F_1\bigg(\frac{\frac{1}{\beta }-1}{\nu }+2,\frac{\mu }{\nu }+2;\frac{\frac{1}{\beta }-1}{\nu
   }+3;-2^{\nu /8} \left(\frac{q^2 \sqrt{\alpha }}{r_c^4}\right)^{\nu /4}\bigg).\nonumber
     \end{align}
     \end{widetext}

    Therefore, in terms of numerical results, the dependencies of the coefficients $a_i$ upon the parameters $q$ and $\beta$ are displayed in Tab. $\ref{Tab2}$. 
    Given that the pressure remains constant during the phase transition, the following satisfies: 
     \begin{equation}
         a+a_1 \,t+a_3 \,t\omega_l+a_5\,\omega_l^3=a+a_1\, t+a_3 \,t\omega_s+a_5\,\omega_s^3.\label{78} 
     \end{equation}
     where $\omega_s$ and $\omega_l$ are the reduced volumes of the small and large black holes, respectively.

     On top of this, Maxwell's equal-area law is quite easily stated as follows:
   \begin{equation}
       \int_{\omega_l}^{\omega_l}\omega\,\frac{\mathrm{d}p}{\mathrm{d}\omega}\,\mathrm{d}\omega=0,\label{79}
   \end{equation}
while taking into account the first derivative, so that\begin{equation}
    \frac{\mathrm{d}p}{\mathrm{d}\omega}=a_3\, t+3\,a_5\,\omega^2.\label{80}
\end{equation}
Using Eqs. \eqref{79} and \eqref{80} yield
\begin{equation}
    a_3\, t(\omega_s^2-\omega_l^2)+ \frac{3}{2}\,a_5(\omega_s^4-\omega_l^4)=0.
\end{equation}
whereupon, and taking into account Eq. \eqref{78}, we may find an explicit link between $\omega_l$ and $\omega_s$ in the following way:
\begin{equation}
    \omega_l=-\omega_s=\sqrt{-\frac{a_3}{a_5}t}\label{82}
\end{equation}
where the argument under the square root function remains positive. A quick look at Eq. \eqref{82}  gives the expected results, namely
\begin{equation}
    \eta=V _l-V_s=V_c(\omega_l-\omega_s)=2 V_c\,\omega_l\propto\sqrt{-t}
\end{equation}
which provides $\lambda=1/2$.

The exponent $\gamma$ is used to assess the critical behavior of the isothermal compressibility $\kappa_T$ stated in terms of
\begin{equation}
    \kappa_T=-\frac{1}{V}\frac{\partial V}{\partial P}\biggr\rvert_{V_c}=-\frac{1}{P_c}\frac{1}{\frac{\partial p}{\partial \omega}}\biggr\rvert_{\omega =0}\propto\, \frac{2r_c}{T_c}t^{-1}
\end{equation}
which results in $\gamma=1$. 

The exponent $\delta$ is responsible for describing the critical behavior of Eq. \eqref{71} on the critical isotherm $T=T_c$. Consequently, the shape of the critical isotherm is set at $t=0$, leading to the following results:
\begin{equation}
    \lvert P-P_c\rvert= P_c  \lvert p-1\rvert= P_c  \,\lvert a_5\, \omega^3\rvert=\frac{P_c\,\lvert a_5\rvert}{V_c^3} \lvert V-V_c\rvert^3
\end{equation}
which proves $\delta=3$.

By considering the previous treatment, it is obviously noted that the four exploited critical exponents are those computed for the case of the charged AdS black holes. This does, in fact, prove that the rational non-linear electrodynamic contribution in the context of Rastall gravity cannot change the critical exponents. Consequently, the present study provides physical similarities along the realm of the $P$-$V$ criticality.   }
 \section{Thermodynamic geometry}
{ Showing the nature of interactions among black hole microstructures is appropriately the main goal of this section. For this purpose, the exploitation of the Ruppeiner geometry permits carrying out such interpretations in that sense. Essentially, the Ruppeiner formalism provides valuable information on the internal dynamics of considered thermodynamic systems by inspecting the sign of the corresponding metric curvature. In practice, the dominant attractive (repulsive) micro-interactions are qualitatively identified from the Ruppeiner negative (positive) scalar curvature. Whereas non-interacting systems like ideal gas are set in whenever the Ruppeiner scalar curvature is vanishing.}

 Therefor, we investigate the thermodynamic geometry in the extended phase space. The previous critical finding is used to perform the associated geothermodynamic analysis. Hence, the probing analysis consists of taking Helmholtz-free energy $F=U-TS$ as an appropriate thermodynamic potential and $X^i=(T, V)$ as a thermodynamic variable that is either intensive or extensive. The first approach, on the other hand, entails dealing with the following thermodynamic metric:
 \begin{equation}\label{r57}
 \mathrm{d}\ell^2=\frac{1}{T}\bigg(-\frac{\partial^2F}{\partial T^2}\mathrm{d}T^2+\frac{\partial^2F}{\partial V^2}\mathrm{d}V^2\bigg),
 \end{equation}
 along with the injection of the differential form of free energy, $\mathrm{d}F=-S\mathrm{d}T+P\mathrm{d}V$, into the corresponding metric, yields
 \begin{equation}\label{r58}
 \mathrm{d}\ell^2=\frac{C_V}{T^2}\mathrm{d}T^2+\frac{(\partial_V P)_T}{T}\mathrm{d}V^2,
 \end{equation}
 where $C_V=T(\partial_T S)_V$ is the specific heat capacity at the constant volume $V$. The requirement for the application of the Ricci scalar is the revelation of a correspondence with the phase transition points of the heat capacity. Indeed, at these points, the Ricci scalar diverges. The use of the related metric in  Eq. \ref{r57} gives an expression for the Ricci scalar as
 \begin{equation}\label{r59}
 R=\frac{\left(\partial_V \, P \right)^2-T^2\left(\partial_{T,V} \, P\right)^2+2T^2\left(\partial_V \, P \right)\left(\partial_{T,T,V} \, P \right) }{2C_V\left(\partial_V \,P\right)^2}.
 \end{equation}
 A quick examination shows that the equation of state depends linearly on the temperature $T$; thus, any derivation up to the second order in $T$ vanishes. The resulting expression for the Ricci scalar is formalized as follows \cite{RP}
 \begin{equation}\label{r60}
 R=\frac{1}{2C_V}\Bigg(1-\bigg(T\frac{\partial_{V,T}\,P}{\partial_V\, P}\bigg)^2\Bigg).
 \end{equation}
 The situation for studying thermodynamic geometry offers in-depth further tools, namely, the extrinsic curvature. Indeed, extrinsic curvature for the metric in Eq. \ref{r57} may be represented as follows \cite{HM}
 \begin{equation}\label{r61}
 K=\frac{1}{2\sqrt{C_V}}\bigg(1-T\frac{\partial_{V,T}\,P}{\partial_V\,P}\bigg).
 \end{equation}
 In the framework of a given Van der Waals fluid, the specific heat capacity at a constant volume is $C_V=3/2\,k_B$. It is straightforward that $C_V=T\left(\partial S/\partial T\right)_{r_+}=0$ and as a consequence, in the following, we will employ the normalized form of curvatures as
 \begin{align}\label{r62}
 R_N&=R\, C_V\\
 K_N&=K\,\sqrt{ C_V},
 \end{align}
{ which can be rewritten explicitly in terms of the parameter space of the black hole system and with the thermodynamic variables $T$ and $V$ as 
\begin{widetext}
    \begin{align}
       K_N& = \frac{\sqrt[3]{6} \pi ^{2/3} \alpha  T \sqrt[3]{V}}{-\frac{(4 \beta -1) \mu  2^{\frac{1}{24} (19 \nu +49)} 3^{\frac{1}{3} (-\nu
   -1)} \pi ^{\frac{\nu +1}{3}} V^{2/3} }{\beta  (\beta  (\nu
   -1)+1)}t_1-\frac{\mu  2^{\frac{1}{24} (19 \nu +49)} \left(\frac{3}{\pi }\right)^{-\frac{\nu }{3}-\frac{1}{3}} V^{2/3}
   \left(\frac{\sqrt{\alpha } q^2}{V^{4/3}}\right)^{\frac{\nu +3}{4}} \left(2^{19 \nu /24} 3^{-\nu /3} \pi ^{\nu /3}
   \left(\frac{\sqrt{\alpha } q^2}{V^{4/3}}\right)^{\nu /4}+1\right)^{-\frac{\mu +\nu }{\nu }}}{\beta }+\alpha\,  t_2}\nonumber\\
   &+\frac{1}{2}\\
  R_N& = \frac{1}{108} \left(54-\frac{216\ 6^{2/3} \pi ^{4/3} \alpha ^2 V^{2/3} (T-4 \beta  T)^2}{(1-4 \beta )^2 \left(t_3+\frac{\mu  2^{\frac{1}{24} (19 \nu +49)} \left(\frac{3}{\pi
   }\right)^{-\frac{\nu }{3}-\frac{1}{3}} V^{2/3} \left(\frac{\sqrt{\alpha } q^2}{V^{4/3}}\right)^{\frac{\nu +3}{4}} \left(2^{19
   \nu /24} 3^{-\nu /3} \pi ^{\nu /3} \left(\frac{\sqrt{\alpha } q^2}{V^{4/3}}\right)^{\nu /4}+1\right)^{-\frac{\mu +\nu }{\nu
   }}}{\beta }+2 \alpha  t_4\right)^2}\right)
    \end{align}
       where
    \begin{align}
        t_1 &= \left(\frac{\sqrt{\alpha } q^2}{V^{4/3}}\right)^{\frac{\nu +3}{4}} \,
   _2F_1\left(\frac{\beta  (\nu -1)+1}{\beta  \nu },\frac{\mu +\nu }{\nu };\frac{\frac{1}{\beta }-1}{\nu }+2;-2^{19 \nu /24}
   \left(\frac{\pi }{3}\right)^{\nu /3} \left(\frac{q^2 \sqrt{\alpha }}{V^{4/3}}\right)^{\nu /4}\right)\nonumber\\
   t_2 &= \left(2-2
   \sqrt[3]{6} \pi ^{2/3} T \sqrt[3]{V}\right)\nonumber\\
   t_3&= \frac{(4 \beta -1)
   \mu  2^{\frac{1}{24} (19 \nu +49)} 3^{\frac{1}{3} (-\nu -1)} \pi ^{\frac{\nu +1}{3}} V^{2/3} \left(\frac{\sqrt{\alpha }
   q^2}{V^{4/3}}\right)^{\frac{\nu +3}{4}} \, _2F_1\left(\frac{\beta  (\nu -1)+1}{\beta  \nu },\frac{\mu +\nu }{\nu
   };\frac{\frac{1}{\beta }-1}{\nu }+2;-2^{19 \nu /24} \left(\frac{\pi }{3}\right)^{\nu /3} \left(\frac{q^2 \sqrt{\alpha
   }}{V^{4/3}}\right)^{\nu /4}\right)}{\beta  (\beta  (\nu -1)+1)}\nonumber\\
   t_4&= \left(\sqrt[3]{6} \pi ^{2/3} T \sqrt[3]{V}-1\right).\nonumber
    \end{align}
\end{widetext} 
}

 \begin{figure*}[ptb!]
    \centering
    \begin{subfigure}[b]{0.5\textwidth}
        \centering
        \includegraphics[scale=0.38]{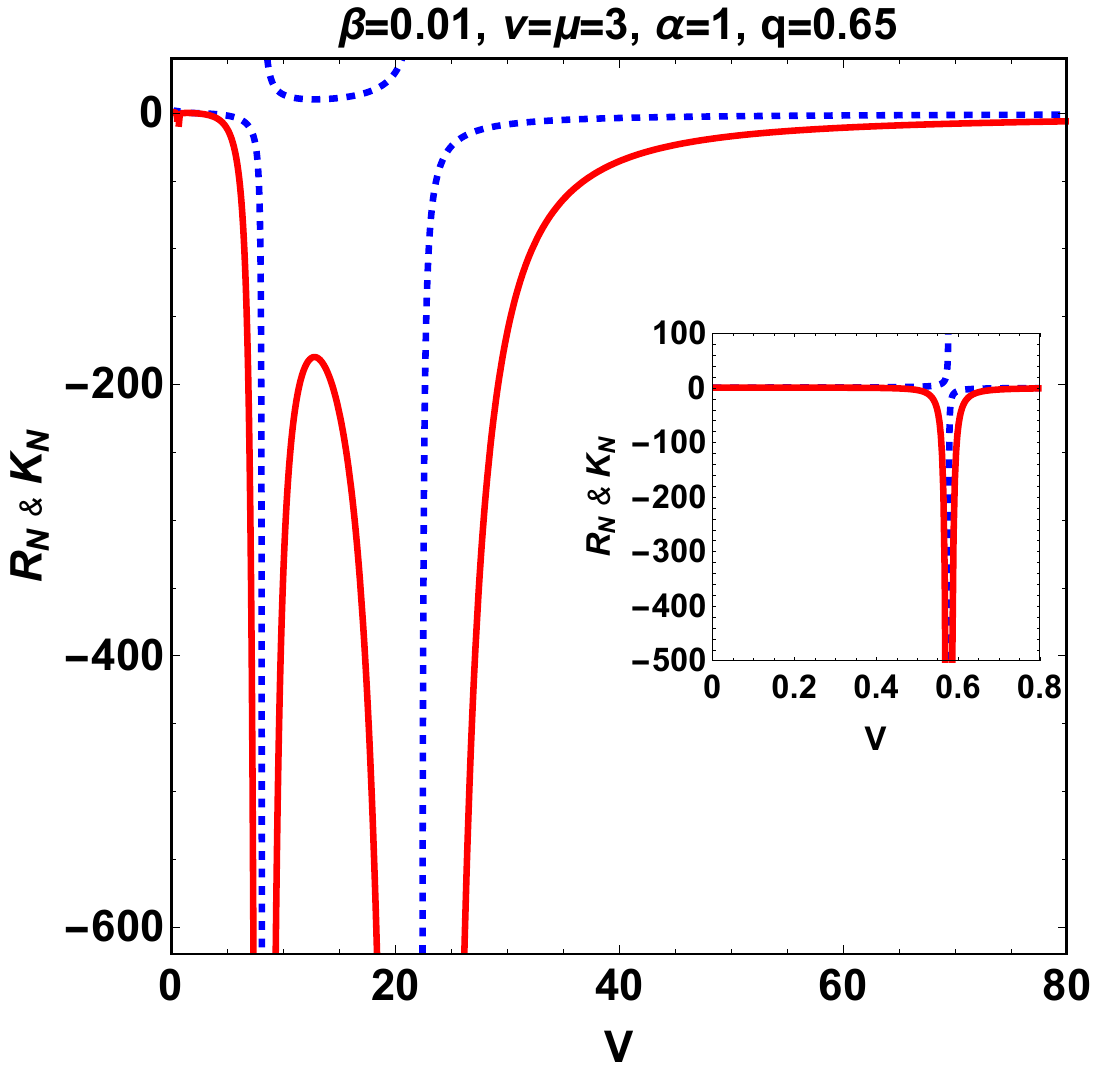}
        \caption{$T<T_c$}
        \label{subfig:wec}
    \end{subfigure}%
    \hfill
    \begin{subfigure}[b]{0.5\textwidth}
        \centering
        \includegraphics[scale=0.38]{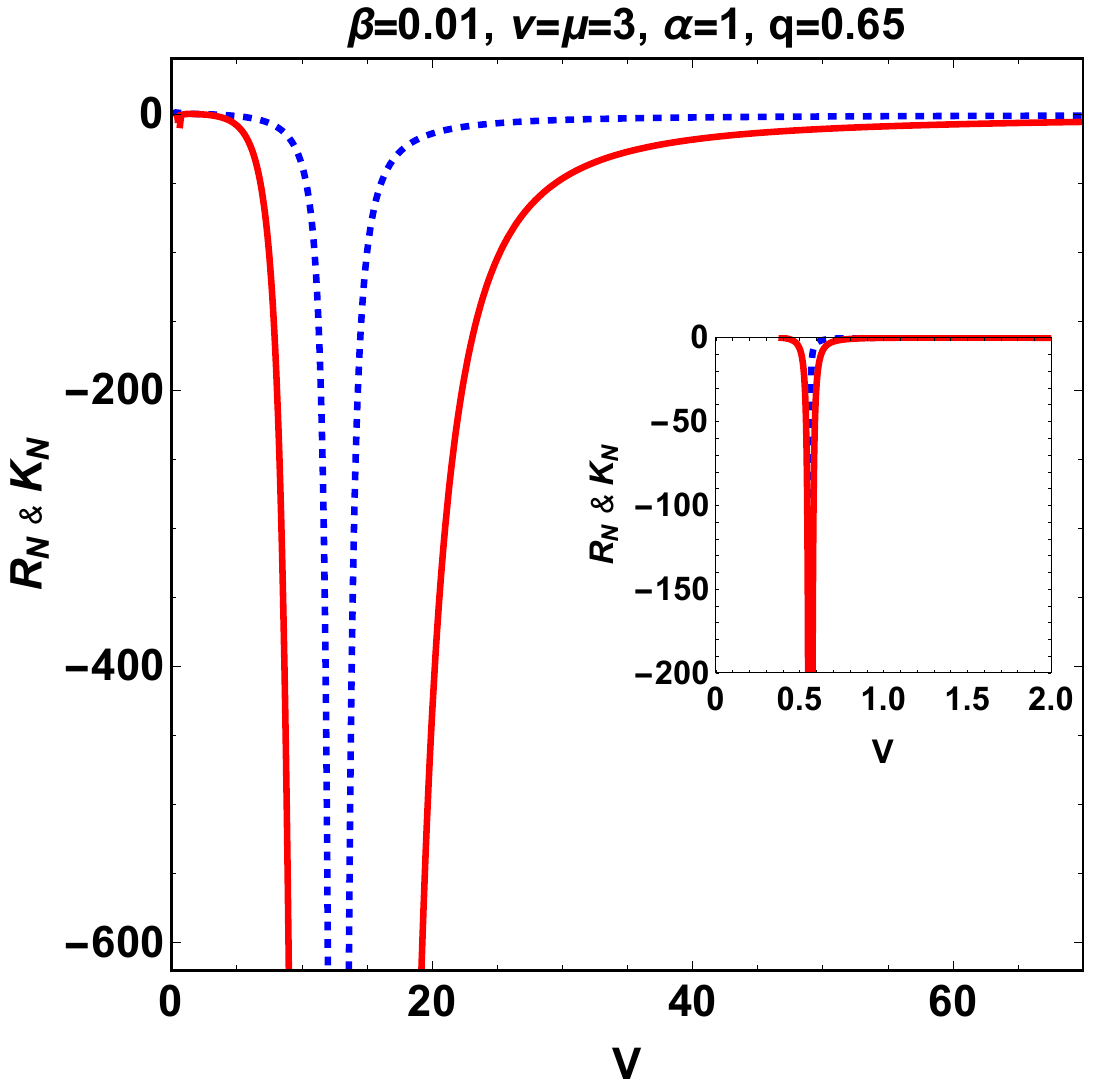}
        \caption{$T=T_c$}
        \label{subfig:nec}
    \end{subfigure}%
    \\
    \begin{subfigure}[b]{0.5\textwidth}
        \centering
        \includegraphics[scale=0.38]{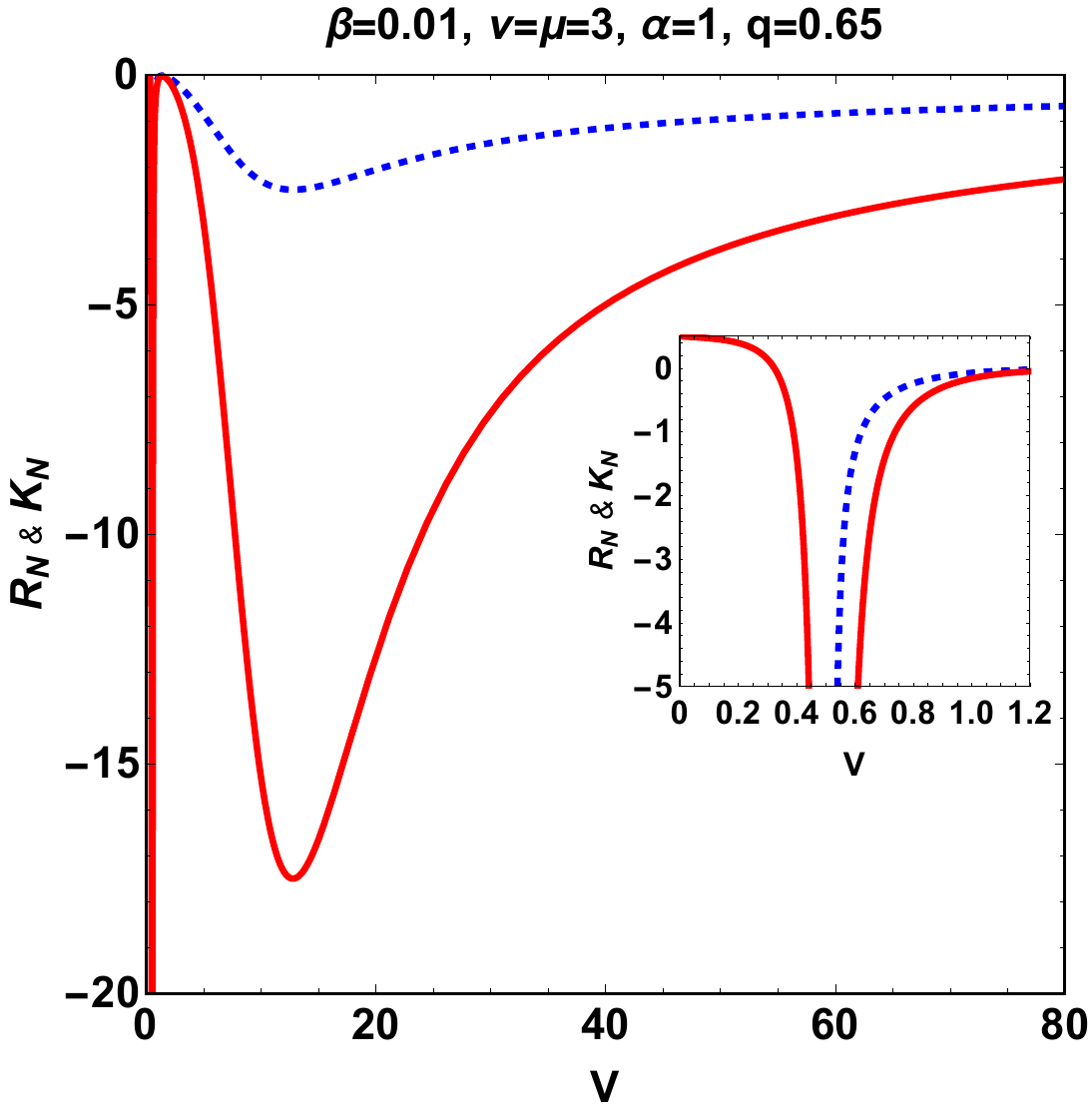}
        \caption{$T>T_c$}
        \label{subfig:wec}%
    \end{subfigure}%
    \caption{Graphs of the normalized thermodynamic Ricci scalar (solid red curve) and extrinsic curvature (dashed blue
curve) as a function of thermodynamic volume $V$.}
    \label{fig7}
\end{figure*}
 The behavior of normalized quantities such as the Ricci scalar and extrinsic curvature is depicted in Fig. \ref{fig7} against the thermodynamic volume. Discovering a particular description for the associated phase transition involving the presence of the Ricci scalar. Concretely, the phase transition takes place at the singularities of the scalar curvature, hence predicting a repulsive interaction between the microscopic black hole molecules in scenarios dealing with $R_N > 0$.
 
 It is remarkable from Fig. \ref{fig7} to note that for $T<T_c$, two critical points are present from which $R_N$ and $K_N$ diverge; the first point is defined at a small $V$ and the second at a large $V$. On the other hand, for $T=T_c$, the two previous divergent points are narrowly reduced to one. Finally, for a temperature higher than the critical temperature, $(T>T_c)$, the singular behavior gets shrunk until it disappears completely. For small values of $V$, a remark takes place concerning the existence of a divergent point despite the temperature changes in all three compared situations.  
  { Additionally, throughout the critical processes, the normalized Ruppeiner curvature scalar turns out to be negative ($R_N < 0)$, indicating the existence of attractive interactions among the black hole microstructures.}
\section{Conclusion and summary}
In this paper, we have constructed a black hole solution with a generalized distribution charged matter source in the Rastall gravity framework. The obtained solution has a number of horizons, depending on the value of the parameter space $\left(M, q, \mu, \nu, \beta, \Lambda\right)$. The considered matter sector, in GR, makes a possible distinction between known kinds of rational non-linear electrodynamics black holes, namely type $(I)$ or type $(II)$, in terms of $\mu$ and $\nu$ parameters, and, at the limit $\nu = 1$ where ($\nu = \mu$), the obtained solution shows a Maxwellian charged matter source. We have studied in the normal phase space the thermodynamic properties, and we have shown analytically and graphically the temperature behavior, heat capacity, and Gibbs free energy. A more detailed analysis concerning global and local stability was conducted by the sign pertinent to heat capacity and Gibbs free energy. As the main goal of this paper is to study the critical behavior of our black hole solution, we have examined the phase structure in which the first law of thermodynamics is extended by adding the $V\,dP$ term, where the cosmological constant $\Lambda$ is considered the pressure $P$.  Furthermore, the critical behavior is investigated along the $P$-$V$ diagram, where at a small value of $r_+$, the isotherms behave like van der Waals behavior, while at a large value of $r_+$, the isotherms turn around and lead again to a region where the pressures are negative. On the same objective, we have treated analytically and graphically the behavior of the Gibbs free energy in the extended phase space, and we have found that there are two critical points with two pressures at which the reentrant phase transition occurs and disappears. For the next step, we have investigated the thermodynamic geometry using the Ruppeiner formalism. In the context of this process, we have demonstrated the normalized quantity for the Ricci scalar as well as the extrinsic curvature. As a result, the normalized Ricci scalar indicates all the possible point-phase transitions of the heat capacity, and the normalized extrinsic curvature has the same sign as the normalized Ricci scalar.

This work comes up with certain issues and will be a topic for revealing the quasinormal modes and shadow behaviors, as well as the deflection angle.
\section*{Acknowledgments}
This work was supported by the Ministry of Science and Higher Education of the Republic of Kazakhstan, Grant
AP14870191.

\bibliographystyle{apsrev}
\end{document}